\documentclass[fleqn,usenatbib]{mnras}

\usepackage{newtxtext,newtxmath}
\usepackage{subcaption} 
\usepackage{xcolor}
\usepackage{orcidlink}

\usepackage[T1]{fontenc}

\DeclareRobustCommand{\VAN}[3]{#2}
\let\VANthebibliography\thebibliography
\def\thebibliography{\DeclareRobustCommand{\VAN}[3]{##3}\VANthebibliography}

\usepackage{graphicx}	
\usepackage{amsmath}	

\title[Astrometry and solar wind]{The impact on astrometry by solar-wind effect in pulsar timing}

\author[K. Liu et al.]{
K. Liu\orcidlink{0000-0003-4453-776},$^{1,2}$\thanks{E-mail: kliu.psr@gmail.com}
A. Parthasarathy,$^{3,2}$
M. Keith,$^{4}$
C. Tiburzi,$^{5}$
S. C. Susarla,$^{6}$
J. Antoniadis\orcidlink{0000-0003-4453-3776},$^{7,2}$
A. Chalumeau\orcidlink{0000-0003-2111-1001},$^{3}$ \newauthor
S. Chen\orcidlink{0000-0002-3118-5963},$^{1}$
I. Cognard\orcidlink{0000-0002-1775-9692},$^8$
A. Golden\orcidlink{0000-0001-8208-4292},$^{6}$
J.-M. Grie{\ss}meier\orcidlink{0000-0003-3362-7996},$^{8,9}$ 
L. Guillemot\orcidlink{0000-0002-9049-8716},$^{9}$
G. H. Janssen\orcidlink{0000-0003-3068-3677},$^{4,10}$  \newauthor
E. F. Keane\orcidlink{0000-0002-4553-655X},$^{11}$
M. Kramer\orcidlink{0000-0002-4175-2271},$^{2,4}$
J. W. McKee\orcidlink{0000-0002-2885-8485},$^{12}$
M. B. Mickaliger\orcidlink{0000-0001-6798-5682},$^{4}$
G. Theureau\orcidlink{0000-0002-3649-276X},$^{8,9,13}$
J. Wang\orcidlink{0000-0003-1933-6498},$^{14}$
\\
$^{1}$Shanghai Astronomical Observatory, Chinese Academy of Sciences, 80 Nandan Road, Shanghai 200030, China \\
$^{2}$Max-Planck-Institut f\"{u}r Radioastronomie, Auf dem H\"{u}gel 69, 53121, Bonn, Germany \\
$^{3}$ASTRON, Netherlands Institute for Radio Astronomy, Oude Hoogeveensedijk 4, 7991 PD, Dwingeloo, The Netherlands\\
$^{4}$Jodrell Bank Centre for Astrophysics, Department of Physics and Astronomy, University of Manchester, Manchester M13 9PL, UK \\
$^{5}$INAF - Osservatorio Astronomico di Cagliari, via della Scienza 5, 09047 Selargius (CA), Italy\\
$^{6}$Physics, School of Natural Sciences, Ollscoil na Gaillimhe --- University of Galway, University Road, Galway, H91 TK33, Ireland \\
$^{7}$FORTH Institute of Astrophysics, N. Plastira 100, 70013, Heraklion, Greece \\
$^{8}$Laboratoire de Physique et Chimie de l'Environnement et de l'Espace, Universit\'e d’Orl\'eans/CNRS, 45071 Orl\'eans Cedex 02, France \\
$^{9}$Observatoire Radioastronomique de Nan\c{c}ay, Observatoire de Paris, Universit\'e PSL, Universit\'e d’Orléans, CNRS, 18330 Nan\c{c}ay, France \\
$^{10}$Department of Astrophysics/IMAPP, Radboud University Nijmegen, P.O. Box 9010, 6500 GL Nijmegen, The Netherlands \\
$^{11}$School of Physics, Trinity College Dublin, College Green, Dublin 2, D02 PN40, Ireland \\
$^{12}$Physics and Astronomy Department, Union College, 807 Union Street, Schenectady, 12308, NY, United States of America \\
$^{13}$Laboratoire Univers et Th\'eories LUTh, Observatoire de Paris, Universit\'e PSL, CNRS, Universit\'e de Paris, 92190 Meudon, France \\
$^{14}$Ruhr-Universit\"at Bochum, Fakult\"at f\"ur Physik und Astronomie, Astronomisches Institut (AIRUB), 44801 Bochum, Germany
}

\date{Accepted XXX. Received YYY; in original form ZZZ}

\pubyear{2024}

\begin{document}

\label{firstpage}
\pagerange{\pageref{firstpage}--\pageref{lastpage}}
\maketitle

\begin{abstract}
Astrometry of pulsars, particularly their distances, serves as a critical input for various astrophysical experiments using pulsars. Pulsar timing is a primary approach for determining a pulsar's position, parallax, and distance. In this paper, we explore the influence of the solar wind on astrometric measurements obtained through pulsar timing, focusing on its potential to affect the accuracy of these parameters. Using both theoretical calculation and mock-data simulations, we demonstrate a significant correlation between the pulsar position, annual parallax and the solar-wind density parameters. This correlation strongly depends on the pulsar's ecliptic latitude. We show that fixing solar-wind density to an arbitrary value in the timing analysis can introduce significant bias in the estimated pulsar position and parallax, and its significance is highly dependent on the ecliptic latitude of the pulsar and the timing precision of the data. For pulsars with favourable ecliptic latitude and timing precision, the astrometric and solar-wind parameters can be measured jointly with other timing parameters using single-frequency data. The parameter correlation can be mitigated by using multi-frequency data, which also significantly improves the measurement precision of these parameters; this is particularly important for pulsars at a medium or high ecliptic latitude. Additionally, for a selection of pulsars we reprocess their EPTA Data Release 2 data to include modelling of solar-wind effect in the timing analysis. This delivers significant measurements of both parallax and solar-wind density, the latter of which are consistent with those obtained at low-frequency band. In the future, combining pulsar timing data at gigahertz and lower frequencies will probably deliver the most robust and precise measurements of astrometry and solar wind properties in pulsar timing.
\end{abstract}

\begin{keywords}
pulsars: general -- astrometry -- solar wind
\end{keywords}


\section{Introduction}
Pulsars are highly magnetised, fast-rotating neutron stars, and at the same time exquisite tools allowing for a variety of astonishing astrophysical studies. In many of these experiments, pulsar astrometry is an essential input. In particular, accurate knowledge of pulsar distances is often a key determinant of the precision of the experiment. For instance, in gravity experiments with binary pulsars, external impact such as the Doppler effect caused by the relative pulsar--solar system motion \citep[Shklovskii effect,][]{shkl70} and gravitational potential in our Galaxy, needs to be corrected to accurately recover the intrinsic variation of the binary orbit \citep[e.g.,][]{lgi+20,ksm+21,gfg+21}. The calculation of these effects requires the knowledge of a pulsar's distance. In a pulsar-timing-array (PTA) experiment to detect gravitational waves (GWs), the precision of pulsar distances is required to be a fraction of the GW wavelength, typically a few parsecs or even smaller, so as to be able to use the `pulsar-term` signal within the PTA response to GWs \citep[e.g.,][]{lwk+11,bps+16}. This capability will significantly enhance the sensitivity of a PTA, especially to continuous GWs from individual sources. Additionally, pulsar astrometry is crucial for understanding the structure of our Galaxy, e.g., for constructing the model of Galactic free electron \citep{cl02,ymw17} and magnetic field distribution \citep[e.g.,][]{hmv+18}. Furthermore, astrometric measurements from pulsar timing can be used to compare the reference frame defined by the JPL planetary ephemeris with that employed by the International Celestial Reference Frame \citep[e.g.,][]{wch+17}, which is fundamental to the comparison and translation of stellar distance measurements based on these two systems \citep{mcc13}. 

Typically, there are two primary ways to precisely measure pulsar astrometry. The first method involves the canonical very-long-baseline-interferometer (VLBI) technique which delivers precise localisation of the pulsar in the sky \citep[e.g.,][]{bbgt02}. Accordingly, pulsar astrometry is achieved by tracking the apparent annual positional variation of the pulsar over a period of one or a number of years. This approach enables precise astrometric measurements for different classes of pulsars including canonical pulsars, millisecond pulsars (MSPs) and magnetars \citep{dvk+16,dgb+19,ddl+20}. For a pulsar at a typical distance of 1\,kpc, the positions and the annual parallax (which is the inverse of the pulsar distance) can be measured with a precision of 0.01--0.1\,milli-arcsecond \citep[mas;][]{dds+23}. 

The other method involves modelling the annual parallax of the pulsar via precision pulsar timing \citep[e.g.,][]{EPTA+2023a}. This timing technique employs the pulsar as a precision clock in its inertial frame. It first measures the times-of-arrival (TOAs) of the pulsar signals at Earth, and then converts the measurements to the pulsar's proper time. The time transformation formula, typically referred to as the timing model, consists of a sequence of time-delay terms that are functions of the pulsar's rotational, astrometric, and orbital parameters. For timing experiments in the radio band, this model also accounts for external propagation effects, such as the frequency-dependent dispersion delay of the pulsar radio signal caused by the ionized interstellar medium. In pulsar timing, the position and annual parallax of the pulsar are modelled through geometric time delays. These delays are caused by the annual orbital motion of the Earth within the solar system and the curvature of the pulsar signal's wavefront due to the finite distance of the pulsar, respectively \citep[e.g.,][]{ehm06}. For a canonical pulsar, its position can be measured with a precision of 10--100 milli-arcsecond \citep[e.g.,][]{wch+17}. Meanwhile, the parallax time delay is only secondary to that caused by the Earth's positional variation (Roemer delay). For a typical parallax of 1\,mas, the parallax time delay amounts to several hundred nanoseconds. As a result, precise parallax measurements are feasible only for pulsars with high precision timing, predominantly MSPs. Typically, the position and parallax of an MSP can be measured with a precision of 10--100 micro-arcsecond ($\upmu$as) \citep[e.g.,][]{EPTA+2023a}.

However, pulsar astrometry derived from timing analysis may be affected by other effects that exhibit similar and partially correlated features in the timing data. Typically, timing parameters are determined simultaneously by fitting the entire timing (and sometimes the noise) model to the dataset. This methodology inherently captures the correlations among various effects present in the timing data. Hence, the measurements of astrometric parameters may be significantly biased if correlated effects are not properly accounted for. For instance, \citet{mcc13} demonstrated how red noise in the timing data can compromise the accuracy of derived astrometric parameters. Their study showed that applying a whitening filter to the data noticeably reduces such biases, leading to more reliable astrometric measurements.

Here, we present a detailed investigation into the impact on astrometric measurements by the solar-wind effect, with a focus on the timing parallax. It is known that the solar wind produces free electrons, altering the local free electron density in the solar system and thus causing an additional frequency-dependent dispersion delay in the radio emissions from pulsars \citep[e.g.,][]{imn+98}. Substantial analyses have been carried out using pulsar timing as a sensitive tool to probe the solar-wind properties \citep[e.g.,][]{tvs+19,stz20,tsb+21}. Typically, in pulsar timing the solar-wind effect is described by a spherical model parameterized by the number density of free electrons at a 1-AU distance from the Sun \citep[$n_{\rm sw}$, e.g., ][]{ehm06}. In many previous studies, the pulsar's position and parallax were fitted in the timing analysis to obtain the pulsar's distance, when $n_{\rm sw}$ was used as a fixed constant input \citep[e.g.,][]{rsc+21,EPTA+2023a}. The fixed values that have been used for the solar wind amplitude vary significantly. For instance, the default value used in the \textsc{tempo} software package is 10\,cm$^{-3}$, while in the \textsc{tempo2} software package it is 4\,cm$^{-3}$. Using the NANOGrav 11-Year dataset, \citet{mca+19} suggested an average value of 7.9 cm$^{-3}$ among all pulsars. However, the solar-wind density at 1-AU distance is seen to be both time and ecliptic latitude dependent \citep{sbt+13,poi+14,tsb+21,sct+24}, so applying the same $n_{\rm sw}$ value to different pulsars and datasets is probably not optimal. Moreover, as we point out later, if a fixed constant $n_{\rm sw}$ is used which significantly differs from the actual time-averaged value, it can introduce a significant bias in other timing parameters that are strongly correlated, such as the astrometric parameters. The correlation between the timing parallax and solar-wind parameters was already briefly discussed in \cite{sns+05}.

The paper is organized as follows. In Section~\ref{sec:maths} we lay out mathematical framework for understanding the correlation between the astrometric and solar-wind effects. In Section~\ref{sec:results} we present a detailed investigation into the impact on astrometric measurements by the solar-wind effect, using mock-data simulations across various parameter spaces. The results from reprocessing the EPTA `DR2full' dataset guided by our findings is shown in Section~\ref{sec:real}. We conclude in Section~\ref{sec:dis} with a brief discussion.  

\section{The correlation between astrometric and solar-wind parameters} \label{sec:maths}
\begin{figure}
\centering
	\includegraphics[scale=0.6]{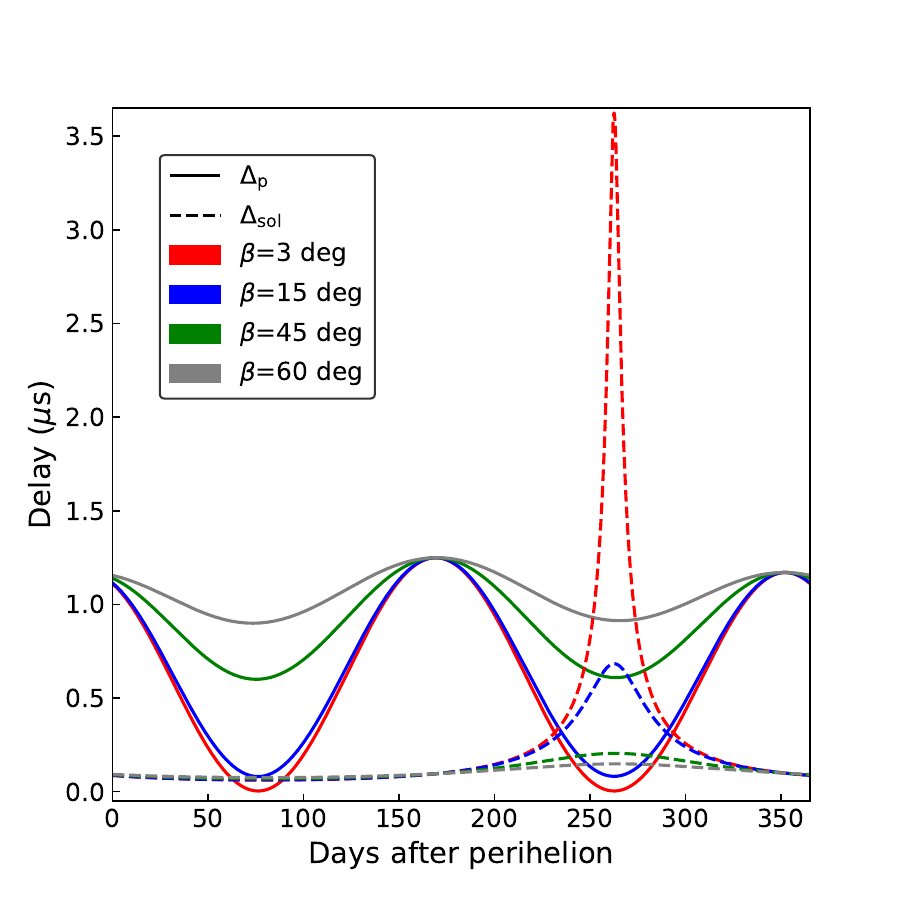}
    \caption{Time delay in pulsar timing caused by annual parallax ($\Delta_{\rm p}$, in solid lines) and solar wind effect ($\Delta_{\rm sol}$, in dashed lines), as a function of days after the perihelion of the Earth. A few values of the pulsar ecliptic latitude $\beta$ were applied, separately, while the ecliptic longitude was set to be $\lambda=0$\,deg.}
    \label{fig:PXvsNE}
\end{figure}

The core of pulsar timing analysis is the time-transformation formula (or the timing model) that is a linear combination of a sequence of time-delay terms. Each term includes a collection of timing parameters to be determined during the timing analysis. The top-level timing model can be expressed as:
\begin{equation} \label{eq:timing_model}
    T_{\rm p} = t_{\rm topo} - \Delta_{\rm R\odot} - \Delta_{\rm p} - \Delta_{\rm sol} - \dots,
\end{equation}
where $T_{\rm p}$ is the pulsar proper time, $t_{\rm topo}$ is the topocentric time. Here $\Delta_{\rm R\odot}$, $\Delta_{\rm p}$, $\Delta_{\rm sol}$ represent the time delay introduced by Earth motion in the solar system (the solar-system Roemer delay), annual parallax of the pulsar and free electron from the solar wind. For simplicity, we only explicitly write the terms directly related to the study in this paper. More details of the pulsar timing model can be found in e.g., \cite{ehm06}. 

The solar-system Roemer delay, $\Delta_{\rm R\odot}$, can be written as
\begin{equation} \label{eq:r_delay}
    \Delta_{\rm R\odot} = - \frac{|\Vec{r}(E)|\cos\theta(E)}{c},
\end{equation}
where $\Vec{r}(E)$ is the Sun--Earth vector, $\theta(E)$ is the angle between Sun--Pulsar and Sun--Earth vectors, $c$ is the speed of light, and $E$ is the eccentric anomaly of the Earth at a given epoch. Following Keplerian's law, $\Vec{r}(E)$ and $\theta(E)$ can be further expressed by:
\begin{eqnarray}
    |\Vec{r}(E)|&=&a(1-e\cos{E}), \\
    \cos{\theta(E)}&=&\cos{\beta}\cos({A_{\rm T}-\lambda)}, 
\end{eqnarray}
where $a$, $e$, $A_{\rm T}$ are the semi-major axis, eccentricity and true anomaly of the Earth orbit, $\lambda$ and $\beta$ are the pulsar longitude and latitude in (geocentric) ecliptic coordinate.

The parallax delay, $\Delta_{\rm p}$, represents the curvature of wavefront of the pulsar signal caused by the finite distance to the pulsar. It is given by \citep[e.g.,][]{ehm06}:
\begin{equation} \label{eq:px_delay}
    \Delta_{\rm p}=\frac{|\Vec{r}(E)|^2\sin^2\theta(E)}{2c}\varpi.
\end{equation}
Here, the parallax $\varpi$ is in unit of arcseconds and $E$ is the eccentric anomaly of the Earth's orbit. 

To first order, the delay introduced by the solar wind, $\Delta_{\rm sol}$, can be approximated by assuming a spherically symmetric free-electron distribution around the Sun. This delay is written as \citep[e.g.,][]{immh98,ehm06}:
\begin{eqnarray} \label{eq:sol_delay}
    \Delta_{\rm sol}= \frac{{\rm DM}_{\odot}}{{\rm D}_{\rm 0}} \cdot \frac{1}{f^2}
    &=&\frac{n_{\rm sw}(1\,\rm AU)^2\theta(E)}{{\rm D}_{\rm 0}|\Vec{r}(E)|\sin \theta(E)} \cdot\frac{1}{f^2},
\end{eqnarray}
where ${\rm D}_{\rm 0}=2.410\times10^{-16} \rm cm^{-3}~pc$ and $f$ is the observing radio frequency. The only model parameter here, $n_{\rm sw}$, is the mean electron density at 1-AU heliocentric radius. Though in reality, the solar wind exhibits complex time variability \citep[e.g.,]{sbt+13,poi+14,stz20,tsb+21,sct+24}, this model is still a good approximation for examining the correlation of solar-wind effect with other timing parameters, and will be used in the subsequent part of the paper.

\begin{figure}
\centering
	\includegraphics[scale=0.5]{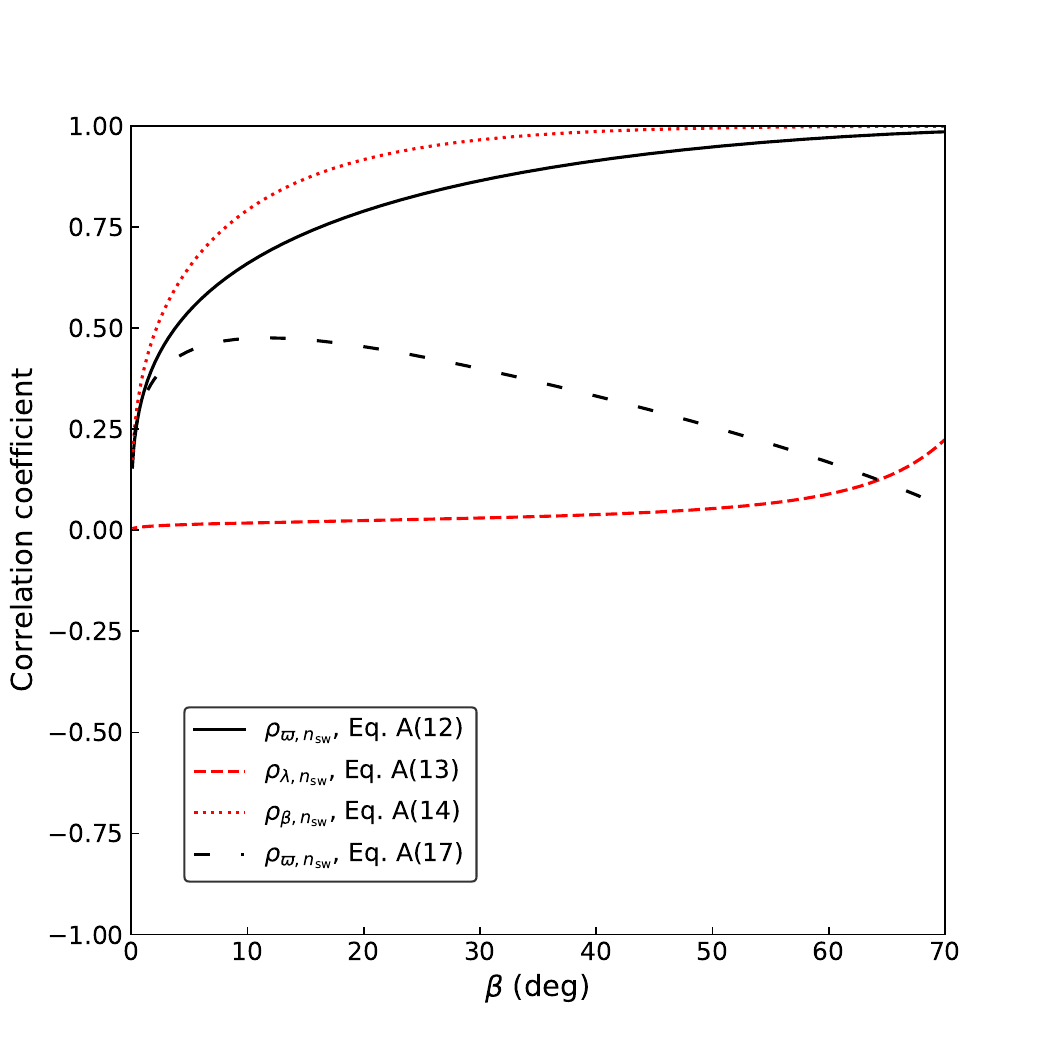}
    \caption{Correlation coefficient of astrometric and solar-wind model parameters, all as a function of ecliptic latitude of the pulsar. The calculations were carried out based on Eq.~(\ref{eq:cc_posi}) and Eq.~(\ref{eq:cc_noposi}). Here the ecliptic longitude $\lambda$ was set to 45\,deg.}
    \label{fig:beta_vs_rho_calcc}
\end{figure}


Clearly, all three delay terms above exhibit annual or bi-annual periodic variation modulated by $\theta(E)$, so some of their model parameters are likely to show some significant correlation. Figure~\ref{fig:PXvsNE} gives the waveform of $\Delta_{\rm p}$ and $\Delta_{\rm sol}$ over the course of a year, for various ecliptic latitudes of the pulsar. It can be seen that the two waveforms follow anti-correlated trends, which is most apparent when the amplitude of the solar-wind effect is around its maximum and the parallax delay reaches a minimum. The correlation coefficient between the waveforms of these two delay terms can be directly calculated via:
\begin{eqnarray}
    \rho_{\Delta_{\rm p},\Delta_{\rm sol}}&=&\frac{\overline{\Delta_{\rm p}\cdot\Delta_{\rm sol}}-\overline{\Delta}_{\rm p}\cdot\overline{\Delta}_{\rm sol}}{\sqrt{\overline{\Delta^2}_{\rm p}-(\overline{\Delta}_{\rm p}})^2\sqrt{\overline{\Delta^2}_{\rm sol}-(\overline{\Delta}_{\rm sol})^2}} \nonumber \\
    &=&\frac{\overline{P\cdot Q}-\overline{P}\cdot\overline{Q}}{\sqrt{\overline{P^2} - (\overline{P})^2} \sqrt{\overline{Q^2} - (\overline{Q})^2}}, \label{eq:cc}
\end{eqnarray}
where
\begin{equation}
    \Delta_{\rm p}=P\cdot\varpi,~~~\Delta_{\rm sol}=Q\cdot n_{\rm sw}, \label{eq:PQ}
\end{equation}
and
\begin{equation}
    P=\frac{|\Vec{r}(E)|^2\sin^2\theta(E)}{2c},~~Q=\frac{(1\,\rm AU)^2\theta(E)}{{\rm D}_{\rm 0}|\Vec{r}(E)|\sin \theta(E)} \cdot\frac{1}{f^2}
\end{equation}
The overline in Eq.~(\ref{eq:cc}) represents time-averaging over the course of a year. For a given function of the eccentric anomaly of the Earth's orbit $E$, its annual average is calculated by
\begin{equation}
    \overline{\mathcal{F}(E)}=\int^{2\pi}_{0}\mathcal{F}(E)(1-e\cos{E})dE.
\end{equation}

As shown in Eq.~(\ref{eq:timing_model}), the timing model is a linear combination of a sequence of delay terms including $\Delta_{\rm p}$, $\Delta_{\rm sol}$. Note that the two model parameters, $\varpi$ and $n_{\rm sw}$, are simply the amplitude of the two terms, $\rho_{\Delta_{\rm p},\Delta_{\rm sol}}$ should directly reflect the correlation of the two parameters in a least-square fit to the timing model. Meanwhile, since $\Delta_{\rm p}$ and $\Delta_{\rm sol}$ follow an anti-correlated trend, an increase in $\Delta_{\rm p}$ (thus in $\varpi$) requires a decrease in $\Delta_{\rm sol}$ (thus increase in $n_{\rm sw}$) to keep $\chi^2$ around its minimum in a least-square fit. This means that $\varpi$ and $n_{\rm sw}$ are positively correlated, i.e., $\rho_{\varpi,n_{\rm sw}}=-\rho_{\Delta_{\rm p},\Delta_{\rm sol}}$. Thus, it is clear from Eq.~(\ref{eq:cc}) that $\rho_{\varpi,n_{\rm sw}}$ does not depend on the value of $n_{\rm sw}$ and $\varpi$ or the observing frequency, but on the waveform of the two delay terms over the course of the year. Both delay terms are functions of the ecliptic latitude $\beta$ of the pulsar, with their modulation increasing as $\beta$ becomes smaller. Hence, $\rho_{\varpi,n_{\rm sw}}$ is likely to exhibit a significant dependency on $\beta$.

Note that the correlation estimate above only takes into account $\Delta_{\rm p}$ and $\Delta_{\rm sol}$. In practice, during the timing analysis, a global fit is usually carried out for all model parameters. Since other parameters, such as the position of the pulsar, could also show a significant correlation with $\varpi$ and $n_{\rm sw}$, the $\rho_{\varpi,n_{\rm sw}}$ calculated from the timing analysis can be significantly different. Nonetheless, the correlation analysis can also be conducted through the covariance matrix in the least-square timing fit \citep{bt76}, and is detailed in Appendix~\ref{sec:app}. Adding the two position parameters, ecliptic longitude $\lambda$ and latitude $\beta$ of the pulsar, leads to the expression shown in Eq.~(\ref{eq:cc_posi}). The derivation including only $\varpi$ and $n_{\rm sw}$ gives Eq.~(\ref{eq:cc_noposi}), exactly the same as in Eq.~(\ref{eq:cc}).

Figure~\ref{fig:beta_vs_rho_calcc} shows the correlation coefficients $\rho_{\varpi,n_{\rm sw}}$, $\rho_{\lambda,n_{\rm sw}}$, $\rho_{\beta,n_{\rm sw}}$ calculated from Eq.~(\ref{eq:cc_posi})--(\ref{eq:cc_posi_beta}) and Eq.~(\ref{eq:cc_noposi}), for different ecliptic latitudes of the pulsar. It can be seen that when the position parameters are also included in the fit, $\rho_{\varpi,n_{\rm sw}}$ is indeed significantly altered; it becomes larger for all $\beta$ values shown in the figure. The correlation between $\beta$ and $n_{\rm sw}$ exhibits a similar trend, and is already significant when $\beta$ is more than a few degree, while for $\lambda$ it is significant only when $\beta$ approaches 70\,deg and above. 



There are several additional factors that could influence the correlations between astrometric and solar-wind parameters when fitting to real data. Firstly, the estimates above assume single-frequency observations, whereas in reality, multi-frequency coverage could mitigate these correlations since the solar-wind effect is chromatic while the others are not. Secondly, the calculations above assume complete sampling of the time-delay features over the course of a year. In practice, however, for low ecliptic latitude pulsars where the solar wind time delay changes rapidly around solar conjunction, a typical cadence (weekly to monthly) of timing observations is likely insufficient to fully resolve the solar-wind feature, which may require daily cadence. Additionally, observing the pulsar when it is very close to the Sun in the sky is difficult. For most single-dish radio telescopes, the reflection and concentration of optical and infrared radiation from the Sun can drastically heat up the receiver, and the convergence of radio emission from the Sun can strongly increase the system temperature of the observation. The resulting lack of coverage around solar conjunction could further impact the correlations between astrometric and solar-wind parameters in the fit to real data. All these aspects will be discussed in the following sections.

\section{Simulation analysis} \label{sec:results}

In this section, we conduct a series of simulation studies to investigate the correlation between astrometric and solar-wind parameters, and their measurability under various circumstances. We utilized the \texttt{fake} plugin in the \textsc{tempo2} software package to generate mock pulsar timing data, i.e., the TOAs. These simulations encompass a broad range of ecliptic latitudes to reflect the actual positional distribution of pulsars in the sky \citep{hmth04}.
In addition to astrometric and solar-wind parameters, the pulsar ephemeris used in the creation of our mock dataset also encompasses a set of other standard timing parameters which are the same in all simulations. This includes the pulsar's spin frequency and its derivative, the dispersion measure (the integrated column density of free electrons between the pulsar and Earth) along with its first and second derivatives--whenever multi-frequency data were simulated--and the pulsar's proper motion in relation to the solar system. All these parameters were fitted simultaneously to the simulated data during the timing analysis. The corresponding covariance matrix of the fit was calculated by the \textsc{tempo2} software package which follows the same framework as used in Appendix~\ref{sec:app}. Note that more details of the simulated data will be presented in each individual subsections below, as they can be different for various simulation studies.

\subsection{Correlation on more realistic occasions} \label{ssec:cc_real}

\begin{figure}
	\includegraphics[scale=0.5]{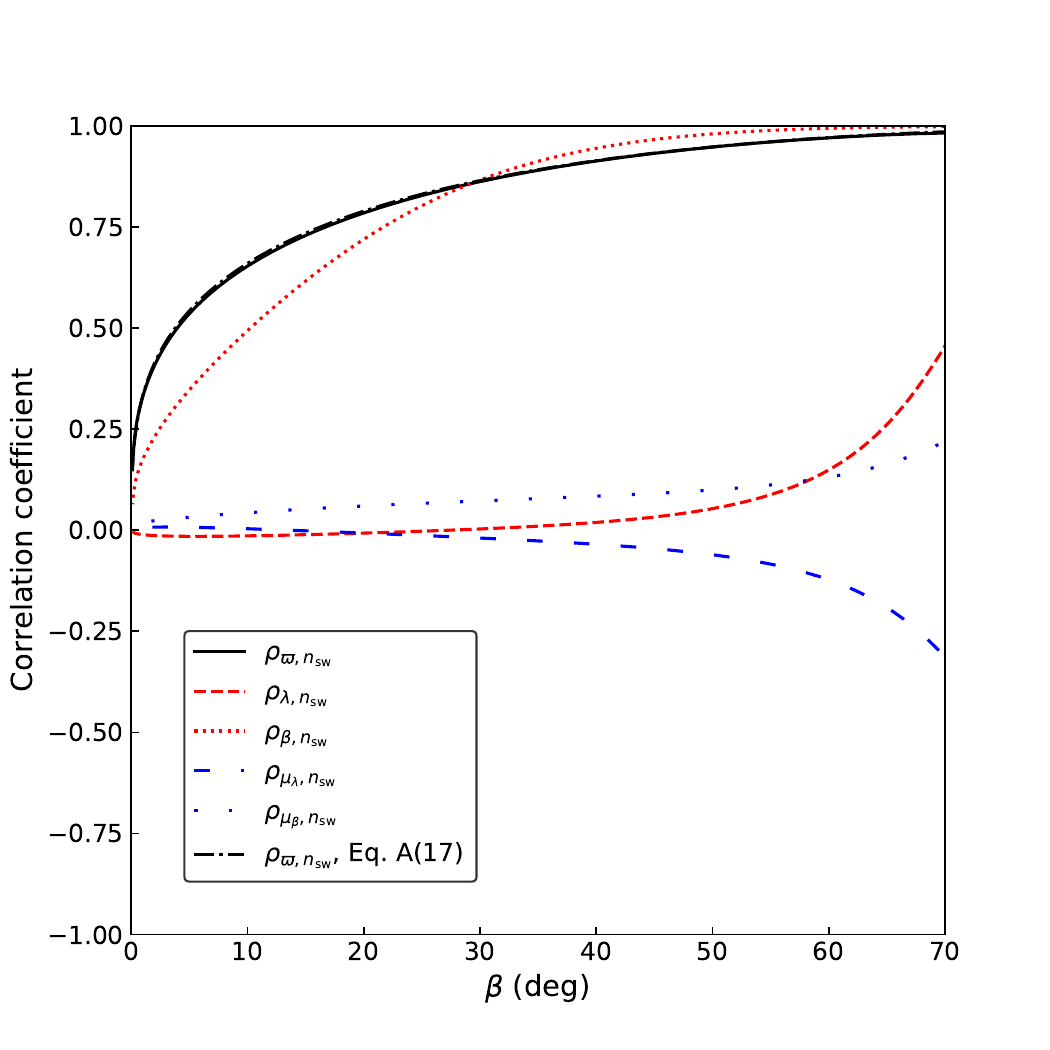}

\vspace{-0.4cm}
 
    \includegraphics[scale=0.5]{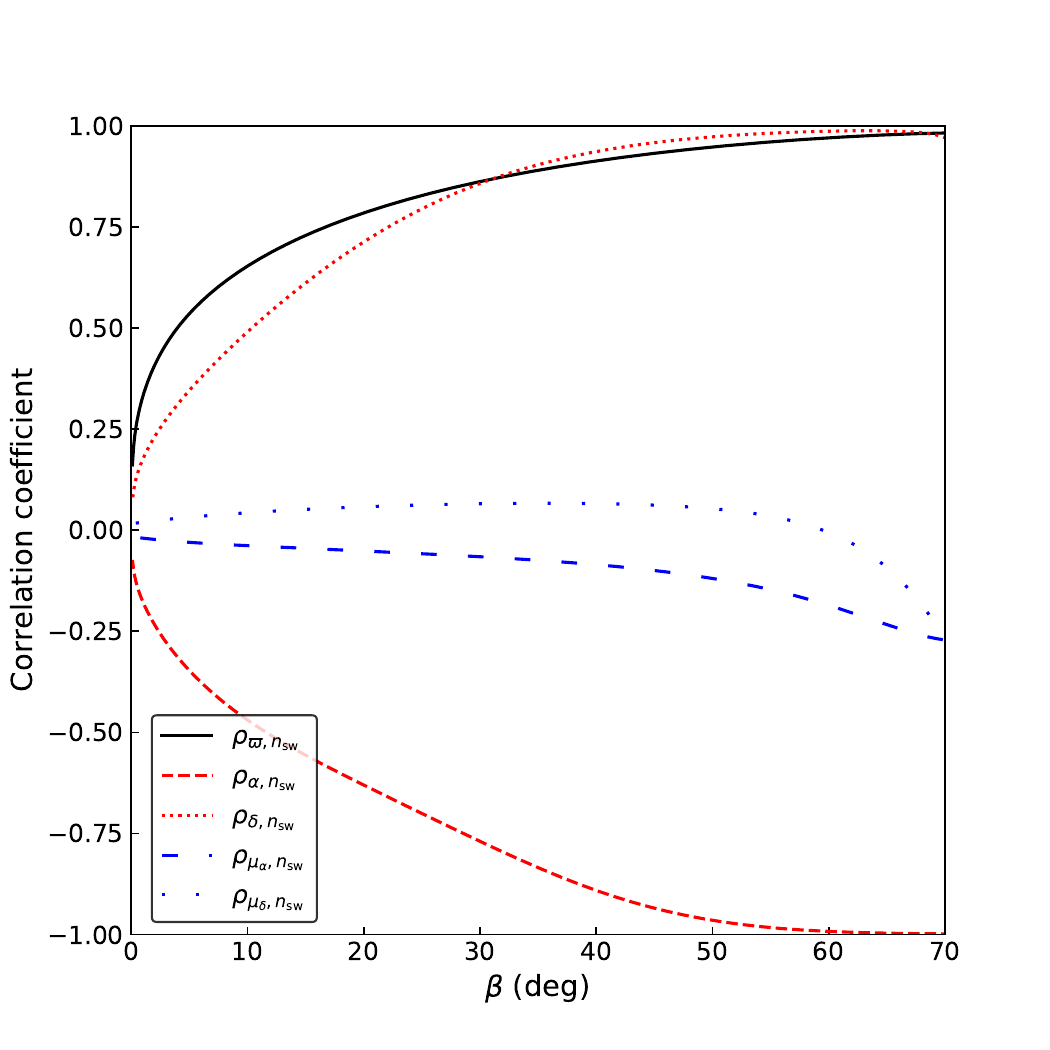}
    \caption{Correlation coefficient of $n_{\rm sw}$ and other astrometric parameters, as a function of ecliptic latitude of the pulsar. The top plot used the ecliptic coordinate system in the timing model, while the bottom employed equatorial coordinate system. The ecliptic longitude was set to 45\,deg in these simulations. }
    \label{fig:beta_vs_rho}
\end{figure}

In a global timing fit alongside other model parameters, the correlations between position, parallax and solar-wind parameters are expected to align closely with the derivations in Eq.~(\ref{eq:cc_posi})--(\ref{eq:cc_posi_beta}), as correlations with other parameters are generally minial. Here, we evaluated the correlations between astrometric and solar-wind parameters under more realistic timing-analysis scenarios. The standard timing model as described above was used to simulated the data at L-band frequency (1.4\,GHz), with high cadence (ten TOAs per day, randomly distributed in time) and precision (1\,ns) to fully capture the sharp feature of solar-wind delay during solar conjunction and accurately estimate the correlation coefficients. The results are presented in Figure~\ref{fig:beta_vs_rho}. It can be seen that $\rho_{\varpi,n_{\rm sw}}$ obtained from the simulation is almost identical as that from Eq.~(\ref{eq:cc_posi}). Regarding the position parameters, $\rho_{\lambda,n_{\rm sw}}$, $\rho_{\beta,n_{\rm sw}}$ also exhibit a very similar trend as in Figure~\ref{fig:beta_vs_rho_calcc}. Meanwhile, it should be noted that here the values of $\rho_{\beta,n_{\rm sw}}$ are generally lower than those from Eq.~(\ref{eq:cc_posi_beta}) at middle and low ecliptic latitudes ($\beta\lesssim50$\,deg). This is mainly because the positional variation in time (i.e., proper motion) mitigates the degeneracy between the position and solar-wind parameters. The proper motion parameters, $\mu_{\rm \lambda}$, $\mu_{\rm \beta}$ start to show significant correlation with $n_{\rm sw}$ only when $\beta$ is approaching 70\,deg. 
In the case where the equatorial coordinate system is used in the timing model, both the position parameters, namely right ascension ($\alpha$) and declination ($\delta$), becomes significantly correlated with $n_{\rm sw}$ when $\beta$ is larger than a few degrees. Meanwhile, $\rho_{\varpi,n_{\rm sw}}$ follows exactly the same trend as in the ecliptic coordinate system, as expected to be independent of the chosen coordinate system. For the remainder of this paper, we will adhere to the ecliptic coordinate system.

As anticipated from Eq.~(\ref{eq:cc_posi})--(\ref{eq:cc_posi_beta}), the correlations in the case of a single observing frequency as above are independent of choice of the frequency and timing precision of the data. Meanwhile, since the solar-wind delay is chromatic while astrometric terms are not, utilizing data with multi-frequency coverage can effectively mitigate the correlation between astrometric and solar-wind parameters. The extent of this mitigation depends on the effective frequency coverage of the data which is determined by the \emph{relative} frequency separation and precision of the data at different frequency bands. For a detailed investigation, we conducted simulations considering a dual-band scenario. In each iteration, the central frequency and timing precision of the reference band were fixed to 1.4\,GHz and 200\,ns, respectively, while those of the other band were set to a different pair of values. Then the corresponding $\rho_{\varpi,n_{\rm sw}}$ were obtained by the timing fit using this dual-band dataset. Note that as indicated by Eq.~(\ref{eq:cc_cov_nopm}), $\rho_{\varpi,n_{\rm sw}}$ is decided by the ratio of the frequency and timing precision at these two bands but not their absolute value, the results of these simulations are also applicable to other choices of the frequency and timing precision of the reference band. 

\begin{figure*}
\centering
\includegraphics[scale=0.385]{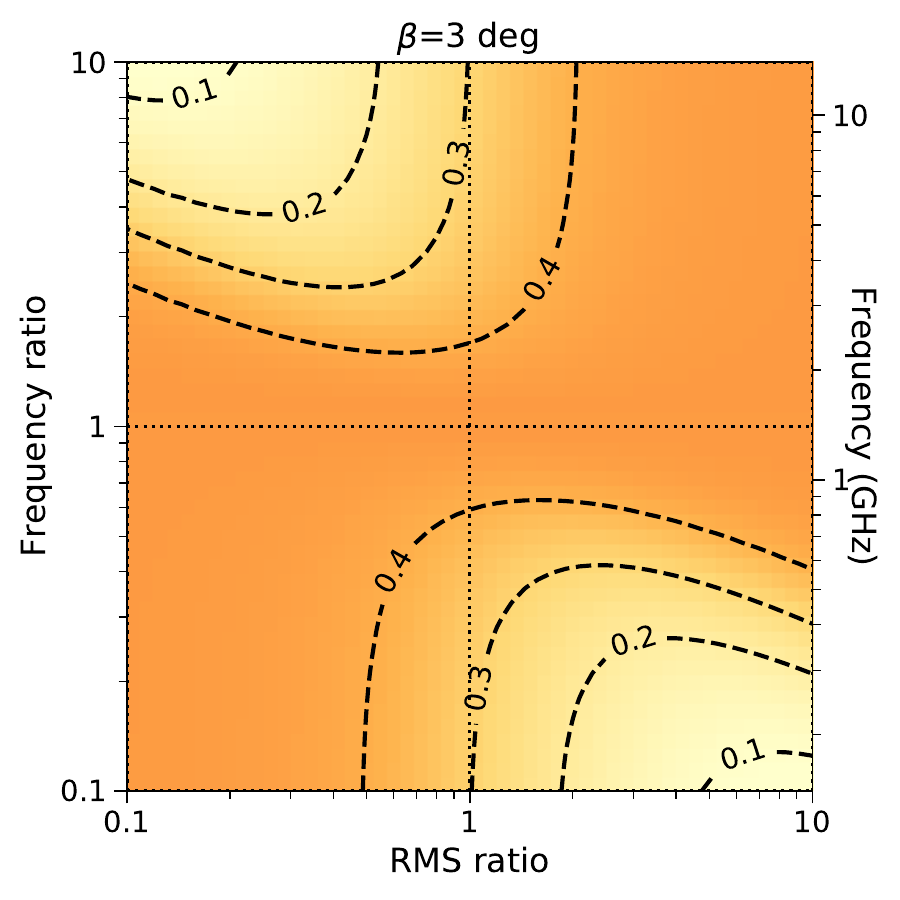}
\includegraphics[scale=0.385]{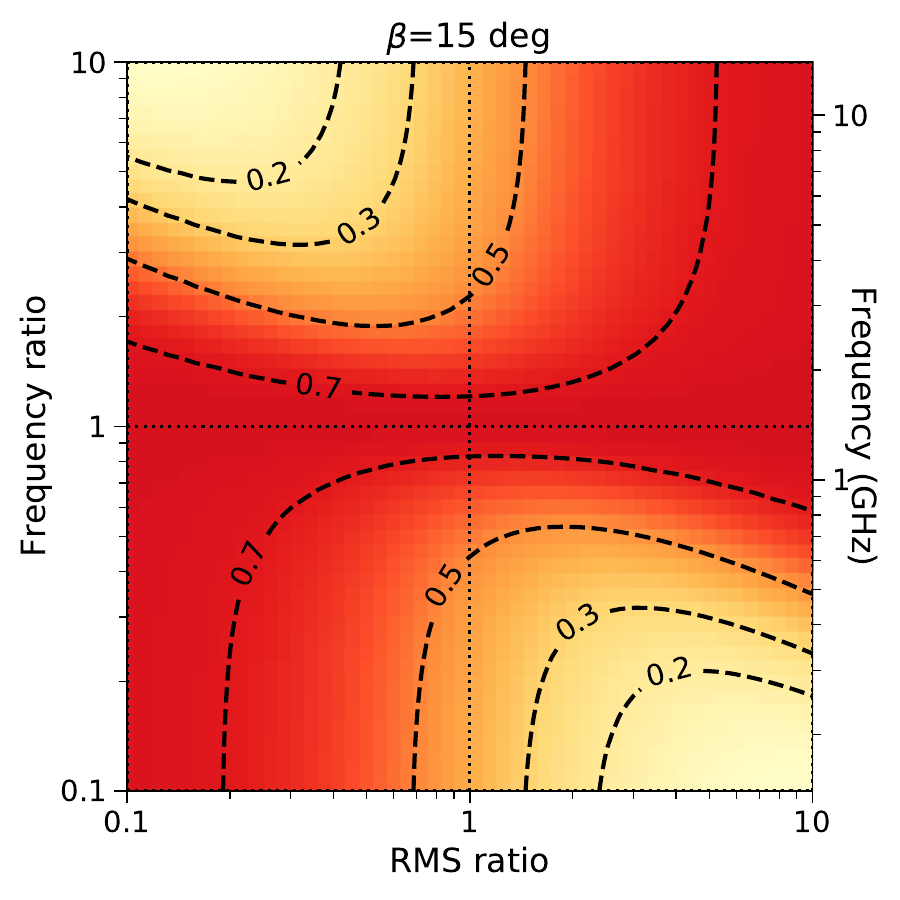}
\includegraphics[scale=0.385]{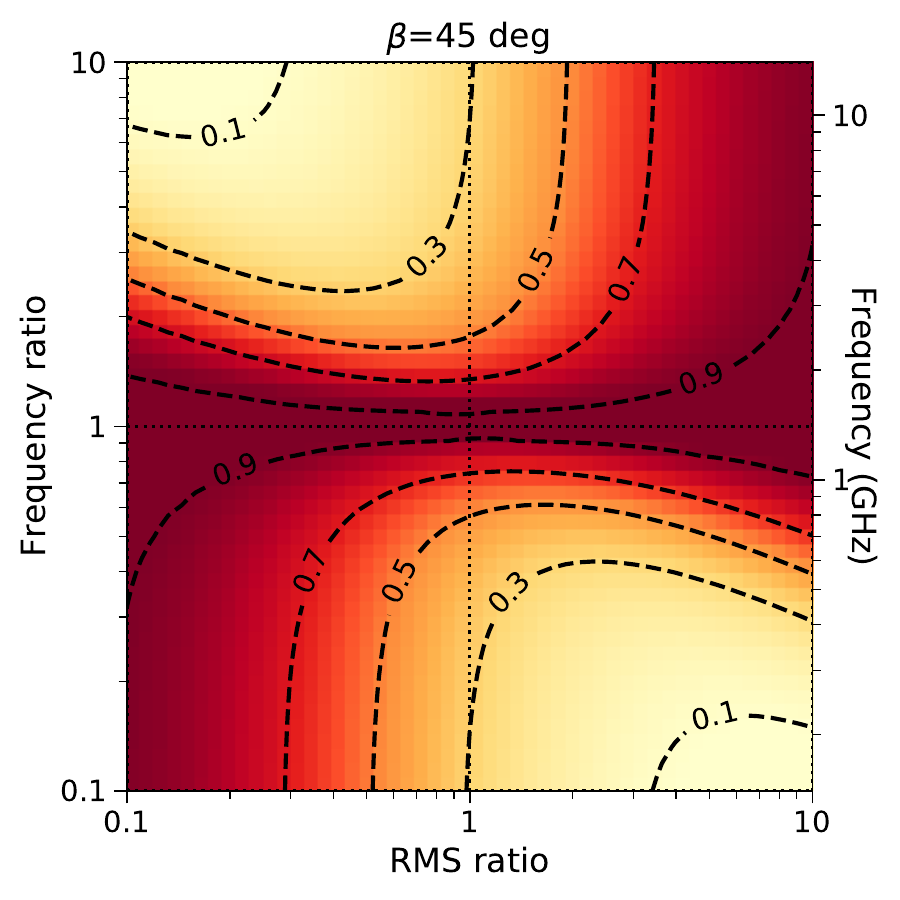}
\vspace{-0.2cm}
    \caption{Correlation coefficient of $n_{\rm sw}$ and $\varpi$ from data with dual-frequency coverage, for various frequency and residual rms ratios of the two bands. These of the reference band were set to 1.4\,GHz and 200\,ns, respectively, in the simulations. The secondary Y-axis shows the frequencies of the alternative band. However, since the correlation coefficient only depends on the frequency ratio of the two bands, these values are directly scalable for other options of the reference frequency. Note that the color scale in all the panels are identical, which deepens as the value increases.}
    \label{fig:fcov_rms_cc}
\end{figure*}

Figure~\ref{fig:fcov_rms_cc} shows the $\rho_{\varpi,n_{\rm sw}}$ for different ratios of the frequency and precision of the alternative band to those of the reference band. The results indicate that for very small $\beta$, the correlation is generally low ($<0.25$) in all cases. However, as the ecliptic latitude increases, effective multi-frequency coverage becomes much more beneficial in mitigating the correlation. For instance, at $\beta=15$\,deg, it can decrease the correlation coefficient from above 0.7 to below 0.3. At 45\,deg, the reduction is even substantial, from over 0.9 down to 0.1. Additionally, for a given frequency ratio, the correlation coefficient is lowest when the high-frequency band has a relatively higher timing precision. This finding suggests that achieving better timing precision at higher frequencies is crucial for effectively mitigating the correlation between the astrometric and solar-wind parameters in multi-frequency pulsar timing analysis. It also aligns with the results from \cite{lbj+14,dvt+20} showing the importance of obtaining higher timing precision in high-frequency observations for DM correction, which requires more dedicated efforts compared to those at low frequencies. 

\subsection{Bias in astrometry for fixed $n_{\rm sw}$}

In timing analysis, it is common to, when applying the spherical model as the correction for the solar-wind delay, give a fixed free-electron density value for the solar wind \citep[e.g.,][]{ehm06,rsc+21,EPTA+2023a}. If this assumption differs from the actual time-averaged value, it would introduce bias in other fitted timing parameters, primarily the astrometric parameters, due to their significant correlations. The bias could be substantial compared to the measurement uncertainty of the parameters, depending on the amplitude of the solar-wind effect primarily decided by the observing frequency and ecliptic latitude of the pulsar compared to the timing precision of the data. To investigate this issue, we conducted a series of simulations with various choices of ecliptic latitude of the pulsar, offset in $n_{\rm sw}$ from the true value, frequency and precision of the TOAs. The solar-wind density was assumed to be time-invariant across the dataset. 

\begin{figure}
\centering
	\includegraphics[scale=0.6]{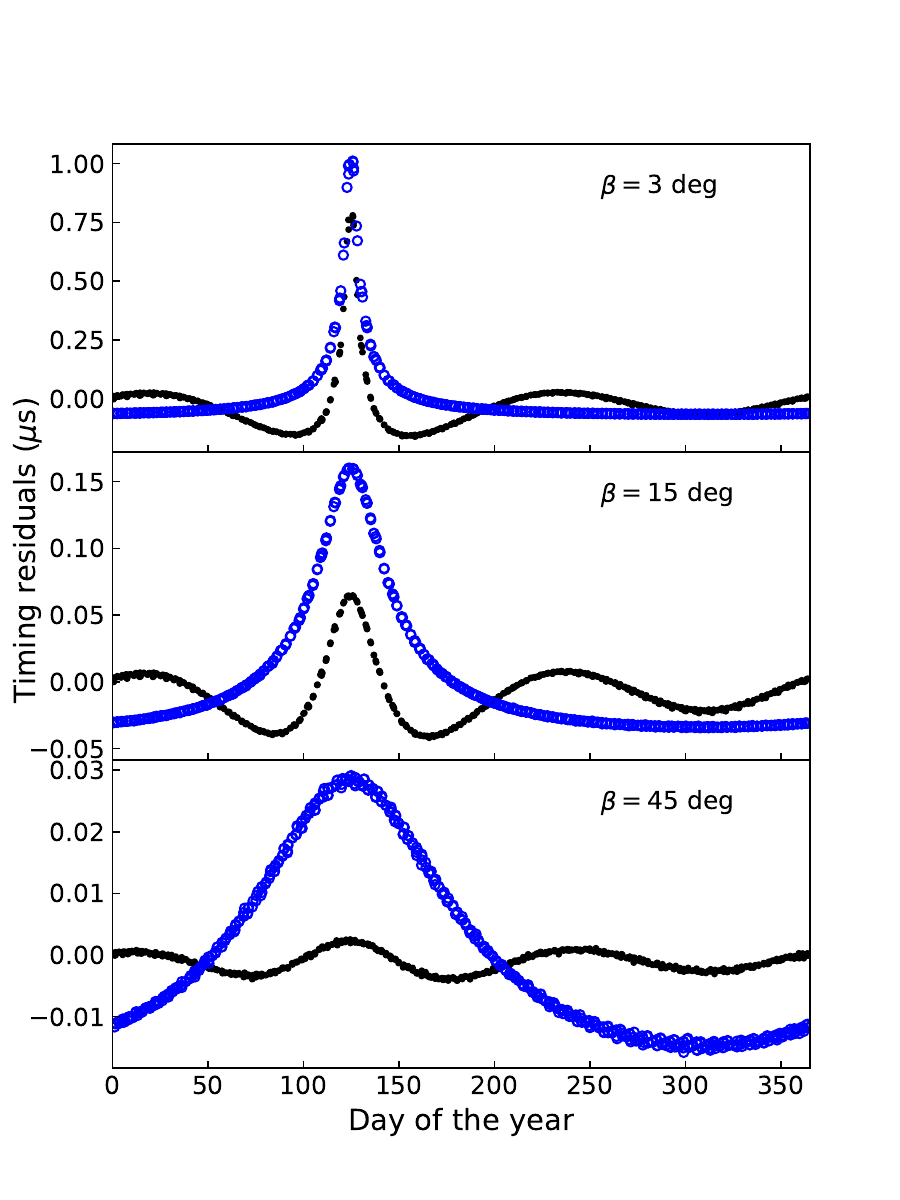}
 \vspace{-1cm}
    \caption{Pre-fit (blue circles) and post-fit (black dots) timing residuals, showing the timing signature of the solar wind from simulations with different ecliptic latitude of the pulsar. Here the TOAs were simulated at 1.4\,GHz and with $n_{\rm sw}=8$\,cm$^{-3}$, while the residuals were created by fixing $n_{\rm sw}$ to 6\,cm$^{-3}$ but fitting all the other timing parameters. The ecliptic longitude here was set to 45\,deg. The TOA precision was assumed to be 1\,ns in this illustration. \label{fig:postfit_res}}
\end{figure}

Figure~\ref{fig:postfit_res} illustrates the absorption of the solar-wind delay feature in the timing data by astrometric parameters. It shows the pre-fit and post-fit timing residuals where $n_{\rm sw}$ was fixed to to 6\,cm$^{-3}$, 2\,cm$^{-3}$ off from the value (8\,cm$^{-3}$) used to simulate the data. It can be seen that for a low ecliptic latitude of $\beta=3$\,deg, the feature is mostly retained after the fit as the correlation is moderate as shown in Figure~\ref{fig:beta_vs_rho}. When $\beta$ increases, the correlation grows rapidly, and thus the remaining feature shrinks after absorption by other timing parameters. For $\beta=15$\,deg, the post-fit residuals are around the 100-ns scale, about half of the pre-fit residuals scale. Note that 100\,ns is approximately the best timing precision achieved so far \citep[e.g.,][]{lgi+20}, suggesting that in this case even with single-frequency data the solar-wind feature could still be measurable with the current best timing precision. Meanwhile, for high ecliptic latitude, e.g., $\beta=45$\,deg, only a small fraction of the solar-wind feature remains after the fit for other timing parameters. This indicates that it would be difficult to break the degeneracy between parallax and solar-wind parameters with single-frequency data in this case. 

\begin{figure}
\centering
	\includegraphics[scale=0.6]{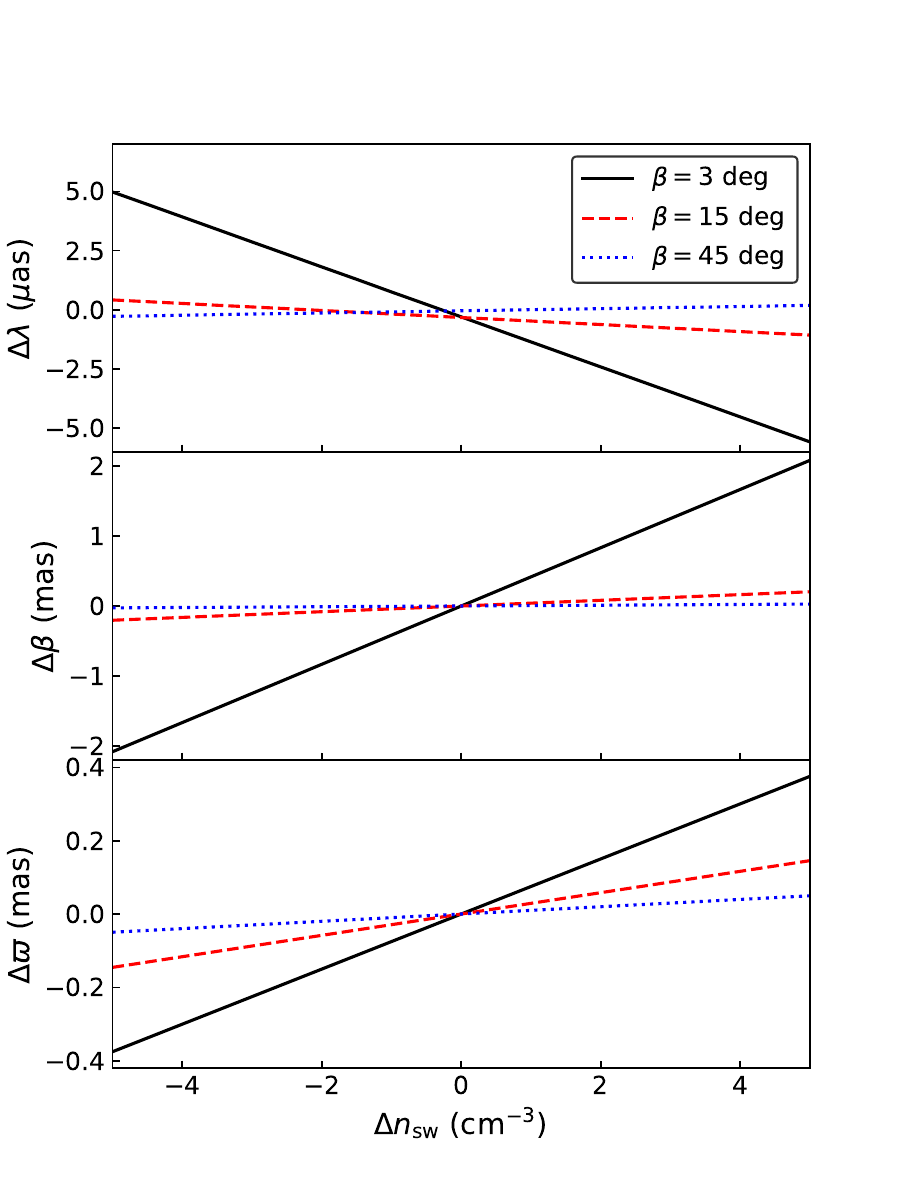}
  \vspace{-1cm}
    \caption{Bias in ecliptic longitude ($\Delta\lambda$), ecliptic latitude ($\Delta\beta$) and parallax ($\Delta\varpi$), induced by the difference in solar-wind density applied in the timing analysis from the actual value ($\Delta n_{\rm sw}$). Here the results are shown for pulsars of different ecliptic latitudes, and the parallax was set to be 1\,mas. The observing frequency of the simulated data was 1.4\,GHz. To accurately estimate these bias values, we applied a very high TOA precision (1\,ns) in the simulation, although these results should be independent of the assumed timing precision of the data as far as the astrometric parameters are significantly measured. The same applies to the results in Figure~\ref{fig:bias_multif}.
    \label{fig:PXbias}}
\end{figure}

Figure~\ref{fig:PXbias} shows the bias in the astrometry parameters caused by the attempt to model the feature caused by various $n_{\rm sw}$ offset values ($\Delta n_{\rm sw}$) with astrometric parameters. It can be seen that as the ecliptic latitude increases, reducing the solar-wind delay, the bias decreases. The bias in ecliptic longitude is generally on the order of 1\,$\upmu$as, smaller than that in ecliptic latitude and parallax, as expected from its lower correlation with $n_{\rm sw}$. The bias in ecliptic latitude is a few orders of magnitude larger, while strongly depending on the value of ecliptic latitude itself. The bias in parallax is generally a few times of 0.1\,mas, which can be critical for some precisely timed pulsars whose timing parallax measurement can be more precise than 0.1\,mas. 

\begin{figure}
\centering
\includegraphics[scale=0.4]{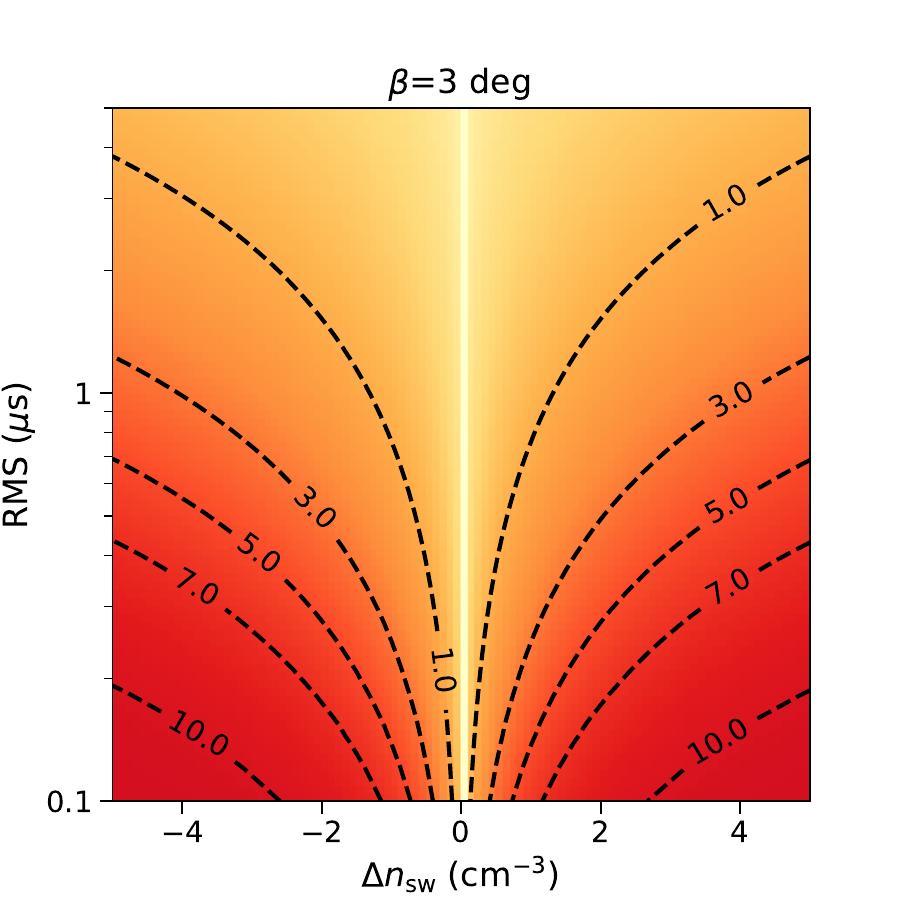}

\vspace{-0.1cm}

\includegraphics[scale=0.4]{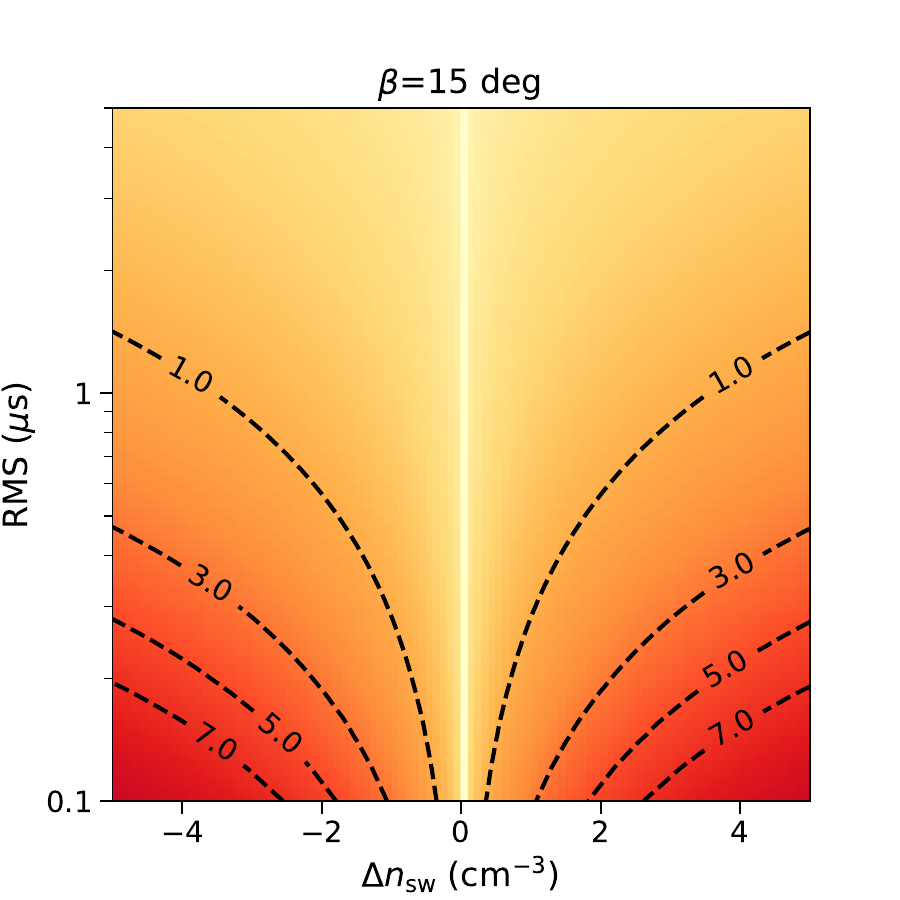}
\includegraphics[scale=0.4]{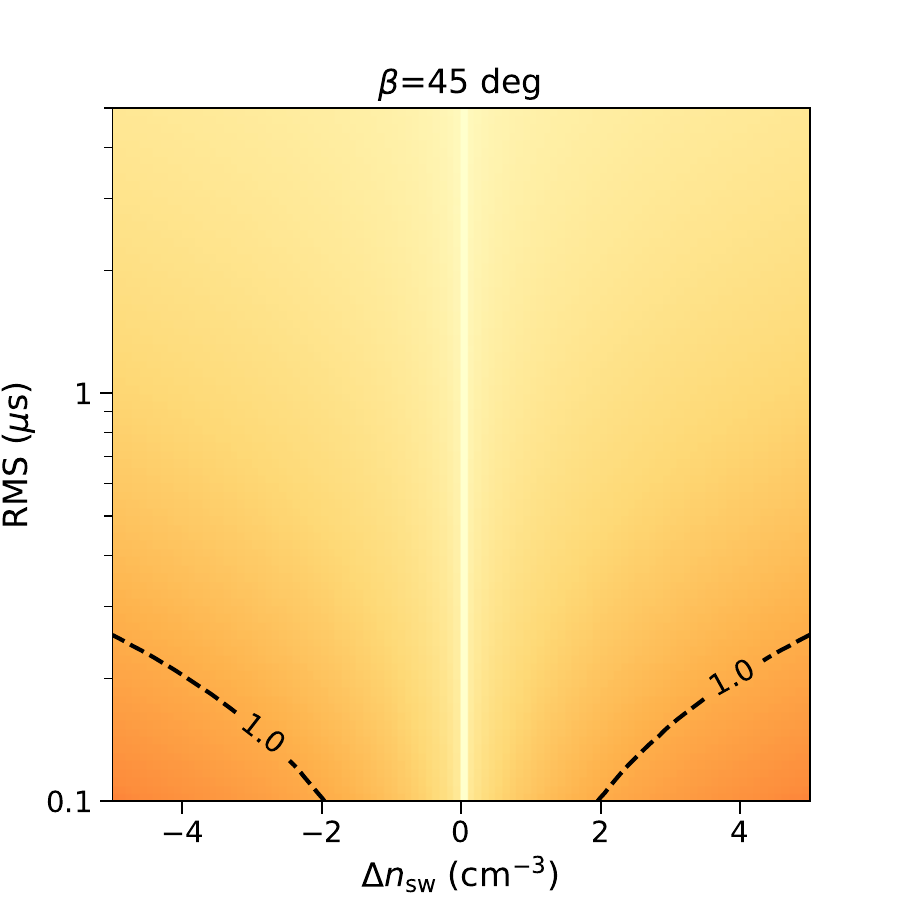}

\vspace{-0.1cm}

    \caption{Heat map showing the absolute value of the bias in fitted parallax normalised by its measurement uncertainty for a few $\beta$ values. The color scales are in all plots are identical so they can be contrasted. Each simulation run here included a five-year-long observation with two TOAs per week. The observing frequency of the simulated dataset was 1.4\,GHz. Note that the color scale in all the panels are identical, which deepens as the value increases. \label{fig:err_ratio_ne_rms}}
\end{figure}

The bias can be either significant or negligible compared to the measurement error of the parallax, strongly depending on the ecliptic latitude of the pulsar, the precision of the dataset, and the deviation of $n_{\rm sw}$. Figure~\ref{fig:err_ratio_ne_rms} shows the normalised bias in parallax by its measurement error for various timing precisions and offsets in $n_{\rm sw}$, across a few values of $\beta$. Here for each simulation run, we generated a five-year observation with two TOAs per week at 1.4\,GHz. It can be seen that when the ecliptic latitude is as low as a few degrees, the bias can be significant, e.g., above 3$\sigma$, for a timing precision around 1\,$\mu$s or even above. At $\beta=15$, it is significant only when the timing precision is a few hundred nanoseconds or better. For $\beta$ at 45\,deg, the bias is generally small compared to the measurement error, unless the timing precision is better than a couple of hundred nanoseconds and the offset in $n_{\rm sw}$ is beyond a few units. In all these cases, the bias significance in ecliptic latitude closely matches that in parallax, as expected from their very similar correlation coefficient shown in Fig.~\ref{fig:beta_vs_rho}. Meanwhile, the bias in ecliptic longitude is always well below 1$\sigma$, owing to its weak correlation with $n_{\rm sw}$.

\begin{figure}
\centering
\includegraphics[scale=0.4]{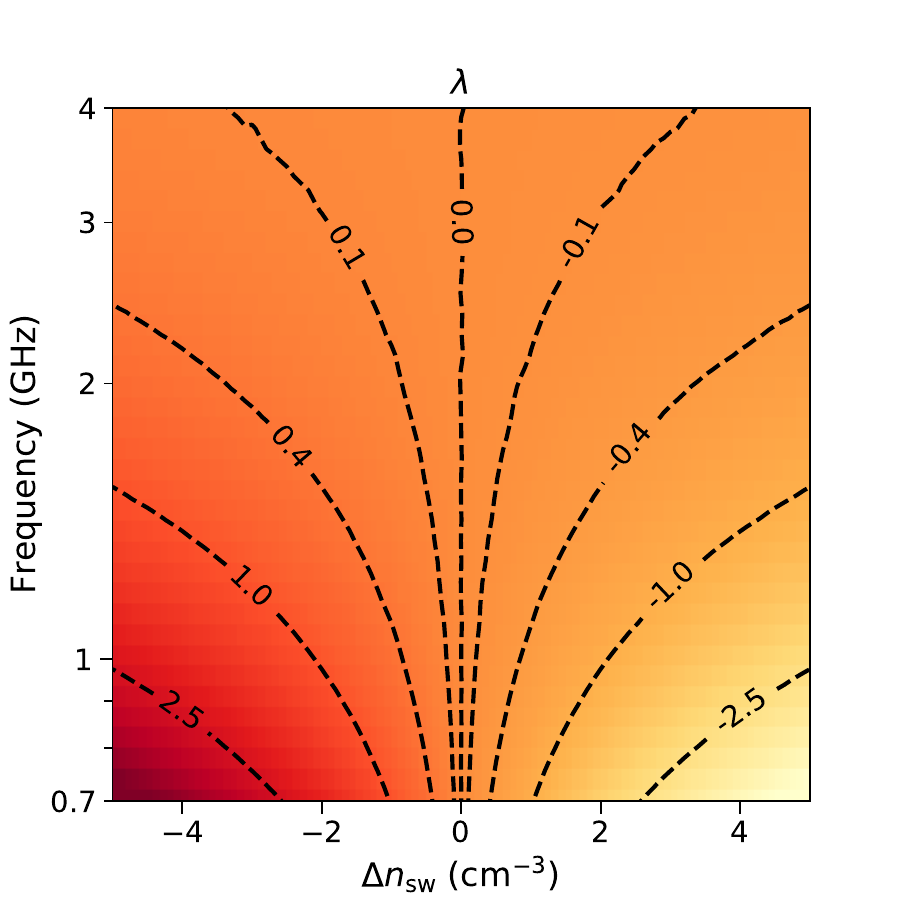}

\vspace{-0.1cm}

 \includegraphics[scale=0.4]{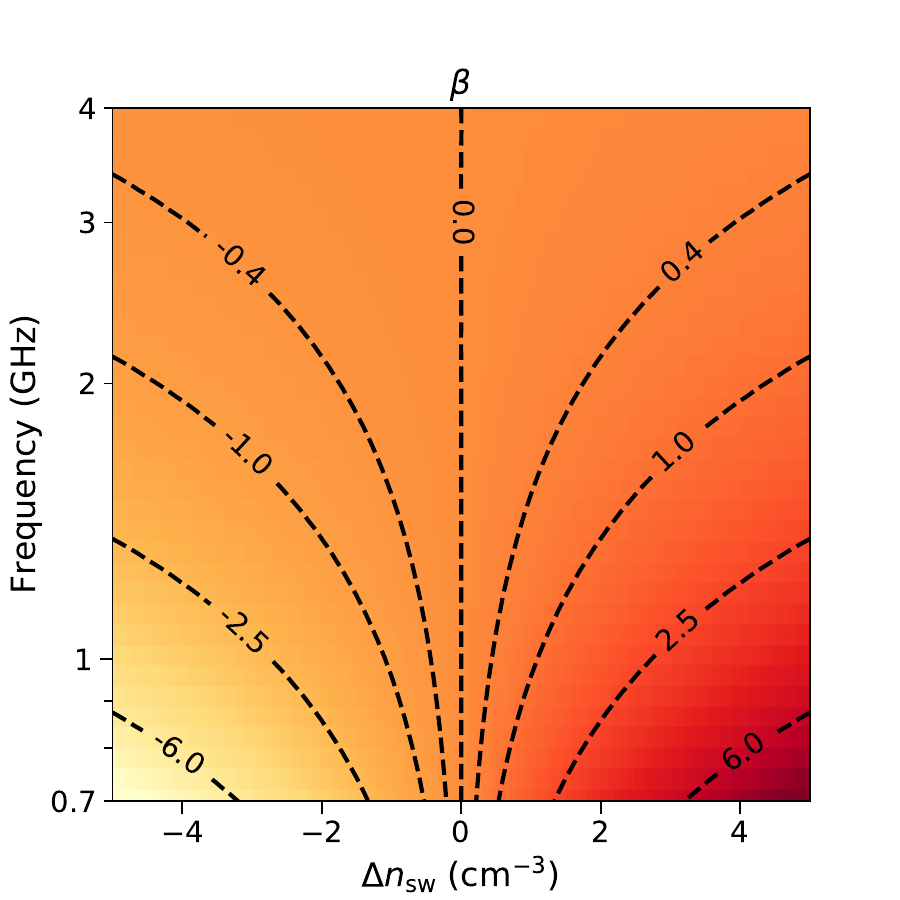}

 \vspace{-0.1cm}
 
 \includegraphics[scale=0.4]{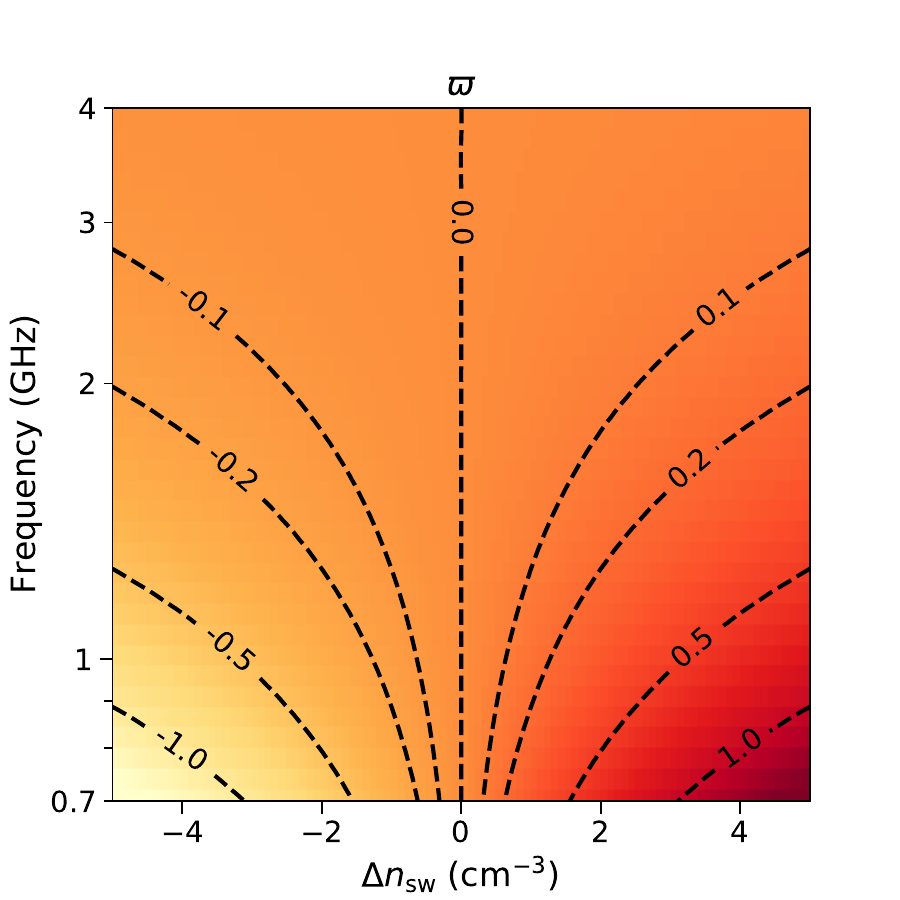}
    \caption{Bias in astrometric parameters for given offset in $n_{\rm sw}$ and observing frequency of the dataset. The biases in $\lambda$ is in unit of $\upmu$as, while the biases in $\beta$ and $\varpi$ are in unit of mas. For these simulations the assumed $\beta$ value is 3\,deg and $\lambda$ is 45\,deg. 
    \label{fig:bias_multif}}
\end{figure}

The scale of the solar wind delay depends on the observing frequency, suggesting that the bias is frequency-dependent as well. For demonstration, in Figure~\ref{fig:bias_multif} we simulated dataset within a wide range of observing frequencies (700\,MHz to 4\,GHz) to inspect how the bias of the astrometric parameter changes. Clearly, for a given offset in $n_{\rm sw}$, the bias in all three astrometric parameters reduces as the frequency increases and the solar-wind effect becomes smaller. Similar to Figure~\ref{fig:err_ratio_ne_rms}, the results again show that for a small ecliptic latitude (e.g., $\beta=3$\,deg) the bias in parallax can be considerable even for observing frequency above 1\,GHz. For observations containing data at multiple frequencies, the overall bias typically falls between the biases at the individual frequencies. The exact value depends on the relative timing precision of each dataset.

Note that bias analysis above may represent a worse-case scenario. In reality, including modelling of DM variation in the timing analysis could be able to absorb some fraction of the solar-wind feature, thereby reducing the bias in astrometric parameters when a fixed $n_{\rm sw}$ is used. This though would highly depend on the assumed model for the DM variation. For instance, the piece-wise DM modelling, commonly referred to as `DMX`, is expected to capture DM variation induced by effects including the solar wind, down to a timescale of a few weeks \citep[e.g.,][]{kcs+13,NG15+a}. Meanwhile, it is not anticipated that a power-law DM noise modelling will fully absorb this feature in the data, since it exhibits an annual periodicity rather than being a stochastic process. Nonetheless, to avoid potential bias in the astrometric parameters, direct modelling of the solar-wind effect should be recommended in the timing analysis, unless the anticipated scale of the effect is clearly well below the sensitivity of the dataset (see Section~\ref{sec:dis} for more discussions). 

\subsection{Fitting for $\varpi$ and $n_{\rm sw}$ in the timing data}

Here we conduct additional simulations to investigate the measurability of $\varpi$ and $n_{\rm sw}$. In these simulations, both parameters are fitted for in the timing analysis, employing the delay model in Eq.~(\ref{eq:px_delay}) and Eq.~(\ref{eq:sol_delay}). Since $\Delta_{\rm p}$ and $\Delta_{\rm sol}$ have a periodic feature with a constant amplitude, the fractional measurement uncertainties of $\varpi$, $n_{\rm sw}$ is expected to scale with respect to the timing precision ($\sigma_{\rm TOA}$), length ($T$), cadence ($C$) of the data as \citep{bt76}:
\begin{eqnarray}
    \frac{\sigma_{\rm \varpi}}{\varpi}\propto\sigma_{\rm TOA}\times\sqrt{\frac{1}{TC}}\times\frac{1}{\varpi} \\
     \frac{\sigma_{n_{\rm sw}}}{n_{\rm sw}}\propto\sigma_{\rm TOA}\times\sqrt{\frac{1}{TC}}\times \frac{1}{n_{\rm sw}}.
\end{eqnarray}
Accordingly, for the simulations in this subsection below, we assume 200\,ns timing precision, 5 years of observations and two TOAs per week without loss of generality. The uncertainties of $\varpi$ and $n_{\rm sw}$ from the simulations can then be scaled with respect to the relations above, to obtain those employing different observational settings. The chosen observing frequency is specified separately in each individual simulation set.

\begin{figure}
\centering
	\includegraphics[scale=0.6]{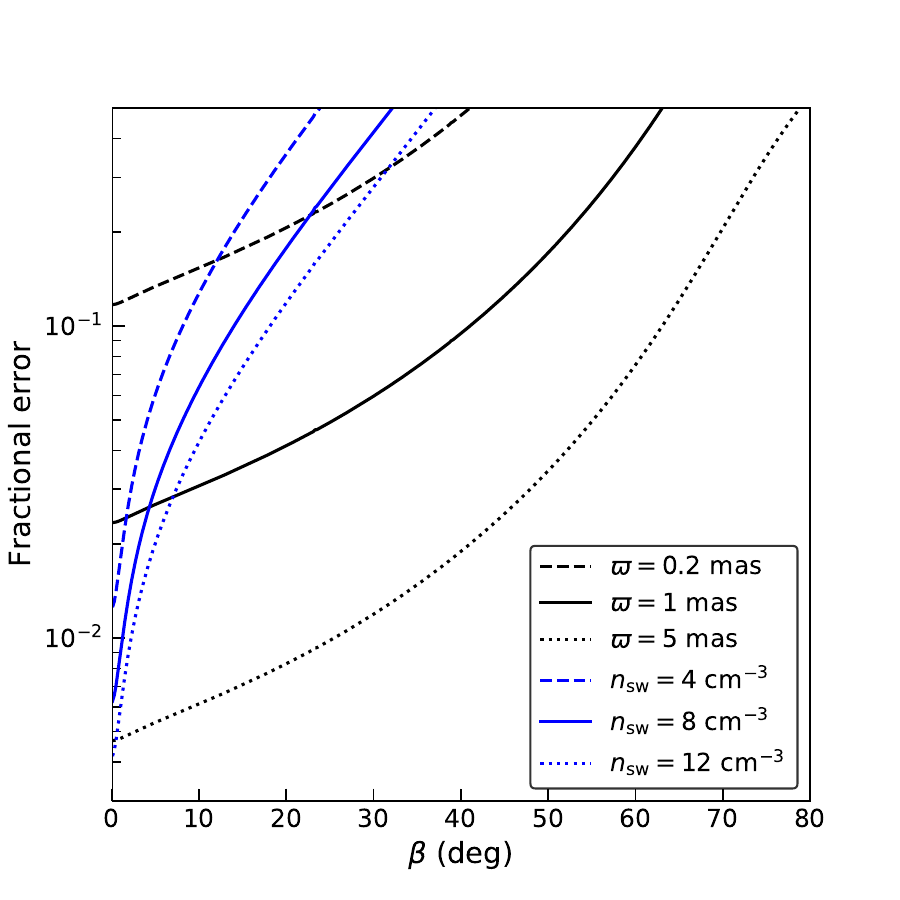}
    \caption{Fractional measurement error of $\varpi$ and $n_{\rm sw}$ as a function of ecliptic latitude $\beta$. The black lines represent the errors of $\varpi$ and the blue lines stand for those of $n_{\rm sw}$. The data were simulated at a frequency of 1.4\,GHz. 
    \label{fig:PXerr}}
\end{figure}

As discussed above, the solar-wind delay has a peculiar deterministic feature in the timing data so is measurable even with single-frequency data. Figure~\ref{fig:PXerr} shows the fractional measurement uncertainties of $\varpi$ and $n_{\rm sw}$ as a function of ecliptic latitude, each for three choices of typical values. Clearly, the measurement uncertainties of both parameters varies strongly with ecliptic latitude and are lower for small $\beta$ values. Notably, in cases where $n_{\rm sw}$ is significantly measured, i.e., with a fractional error less than 20\%, parallax is almost always well measured as well. For $\beta\gtrsim30$\,deg $n_{\rm sw}$ is likely not well constrained, while it is still possible to measure $\varpi$ with a good precision. This is line with the finding by \cite{tvs+19}, that it would be difficult for observations at canonical timing frequencies ($\sim1$\,GHz) to measure solar-wind effect for pulsars at moderate and ecliptic latitudes. 

\begin{figure}
\centering
	\includegraphics[scale=0.45]{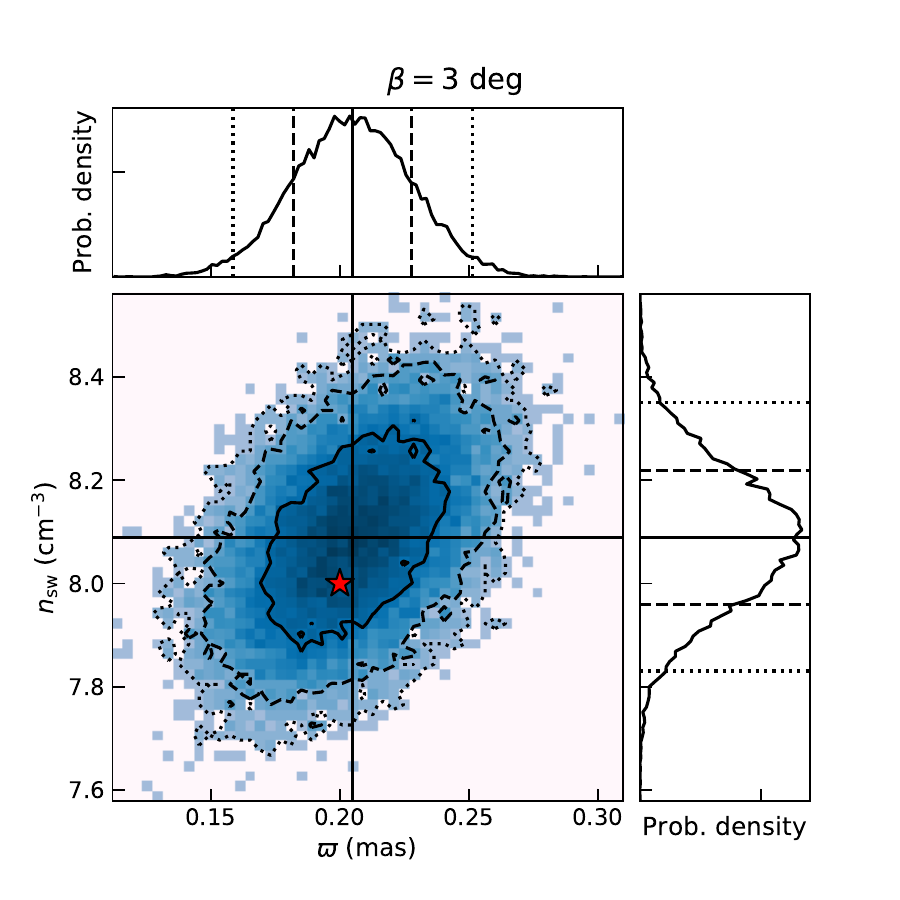}

\vspace{-0.3cm}
 
    \includegraphics[scale=0.45]{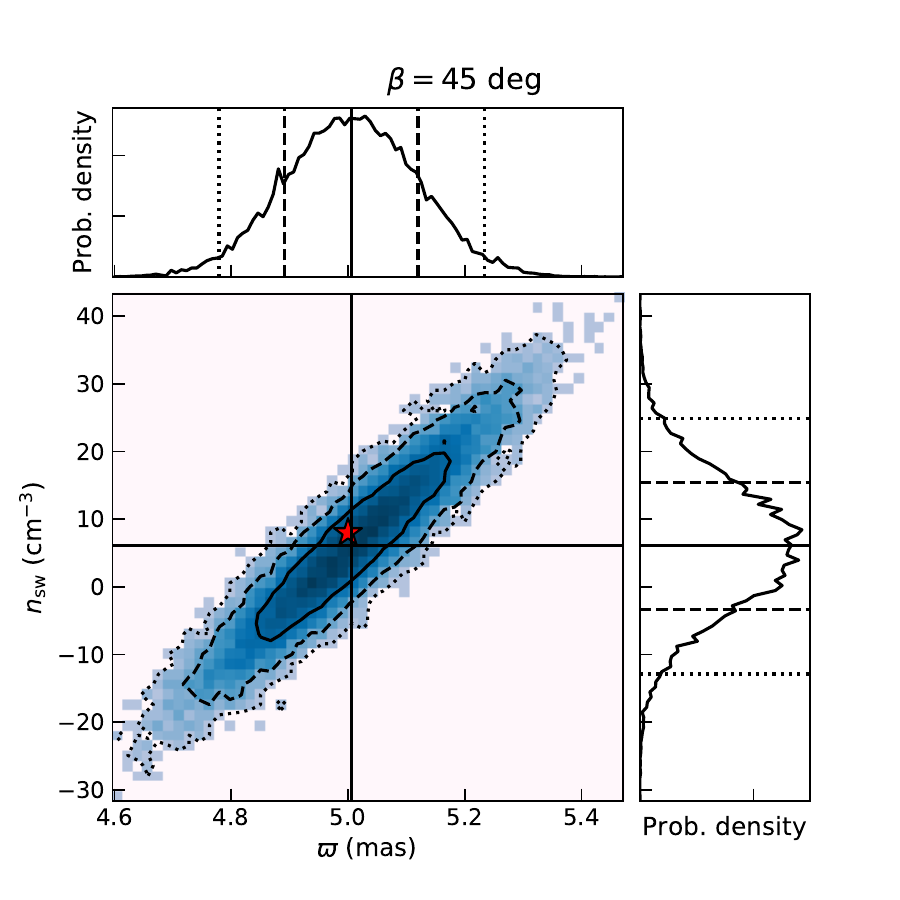}

    \vspace{-0.3cm}
 
    \includegraphics[scale=0.45]{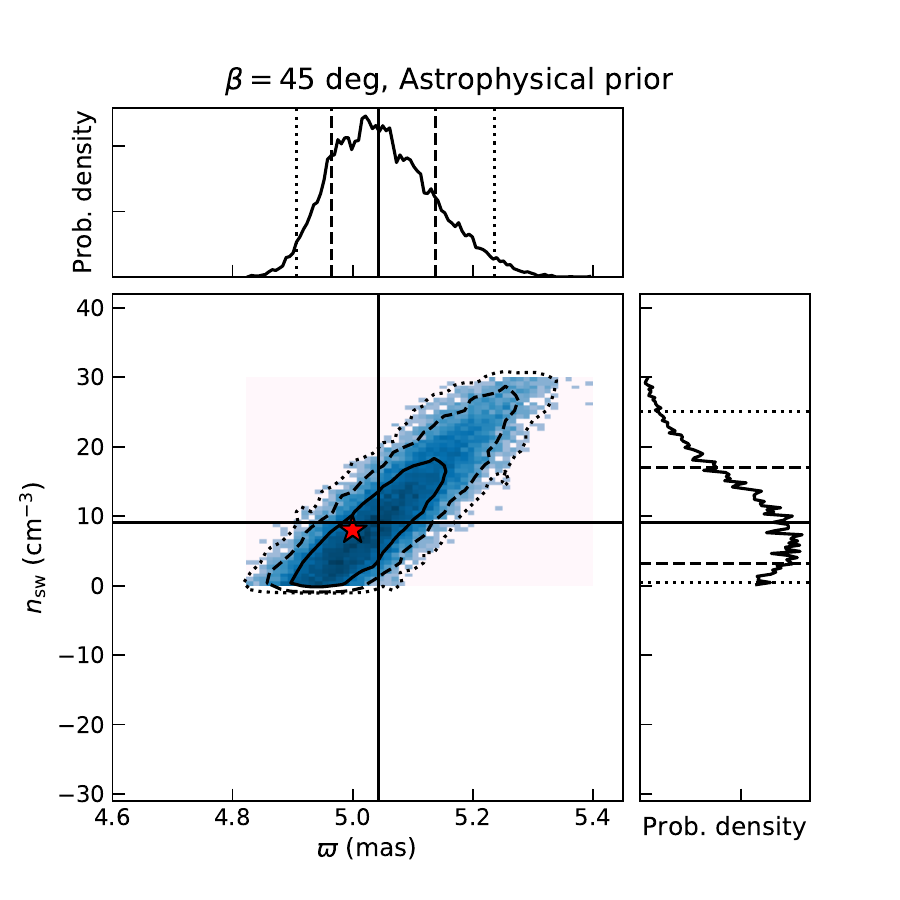}

\vspace{-0.1cm}
    
    \caption{Posterior distribution of $\varpi$ and $n_{\rm sw}$ measured from Bayesian timing analysis of simulated data. The red stars represent the used $n_{\rm sw}$ and $\varpi$ to simulate the data; these are $\beta=3$\,deg, $\varpi=0.2$\,mas, $n_{\rm sw}=8$\,cm$^{-3}$ for the upper panel, and $\beta=45$\,deg, $\varpi=5$\,mas, $n_{\rm sw}=8$\,cm$^{-3}$ for the middle and lower panel. For the analysis in the upper and middle panel, the prior range of $n_{\rm sw}$ was chosen to be $[\overline{n}_{\rm sw}-20\sigma_{\rm 0}, \overline{n}_{\rm sw}+20\sigma_{\rm 0}]$, where $\overline{n}_{\rm sw}$ is the measurement value from the least-square fit and $\sigma_{\rm 0}$ is its uncertainty. In the case of the middle panel, the prior starts from well below zero to nearly -200\,cm$^{-3}$. The prior range used in the analysis of the lower panel is $[0, 30]$\,cm$^{-3}$, well including all known measurements of $n_{\rm sw}$ by far. The ranges of the X and Y axes in the middle and lower panel were set to the same for comparison purpose. The solid, dashed and dotted lines represent $1\sigma$, $2\sigma$, $3\sigma$ contours, respectively. \label{fig:PXNE_3post}}
\end{figure}

For better demonstration of the fitting result when $\varpi$ is well measured and $n_{\rm sw}$ is not, Figure~\ref{fig:PXNE_3post} shows the joint posterior distribution of $\varpi$ and $n_{\rm sw}$ obtained from a Bayesian timing analysis using the \textsc{temponest} software package. Here two cases were considered: one $\beta=3$\,deg, $\varpi=0.2$\,mas, and the other with $\beta=45$\,deg, $\varpi=5$\,mas, respectively. The first case serves as a reference where both $n_{\rm sw}$ and $\varpi$ are well measured. In the second case, $n_{\rm sw}$ is not measured significantly and the constraint of its value placed by the Bayesian timing analysis more depends on the chosen prior range. Nonetheless, the global fit including both $n_{\rm sw}$ and $\varpi$ can still deliver a precise measurement of $\varpi$, even with a conservative choice of prior range for $n_{\rm sw}$ (see the caption of Figure~\ref{fig:PXNE_3post}). Using a more informative (and astrophysical) prior slightly improves the measurement precision of $\varpi$. Although in this scenario applying a fixed $n_{\rm sw}$ in the timing analysis may become feasible as it may introduce only a minimal bias in the measured parallax as shown in e.g., Figure~\ref{fig:PXbias}, conducting the joint fit analysis would still provide useful information for deciding the value of the fixed $n_{\rm sw}$. 

\begin{figure}
\centering

 
    \includegraphics[scale=0.55]{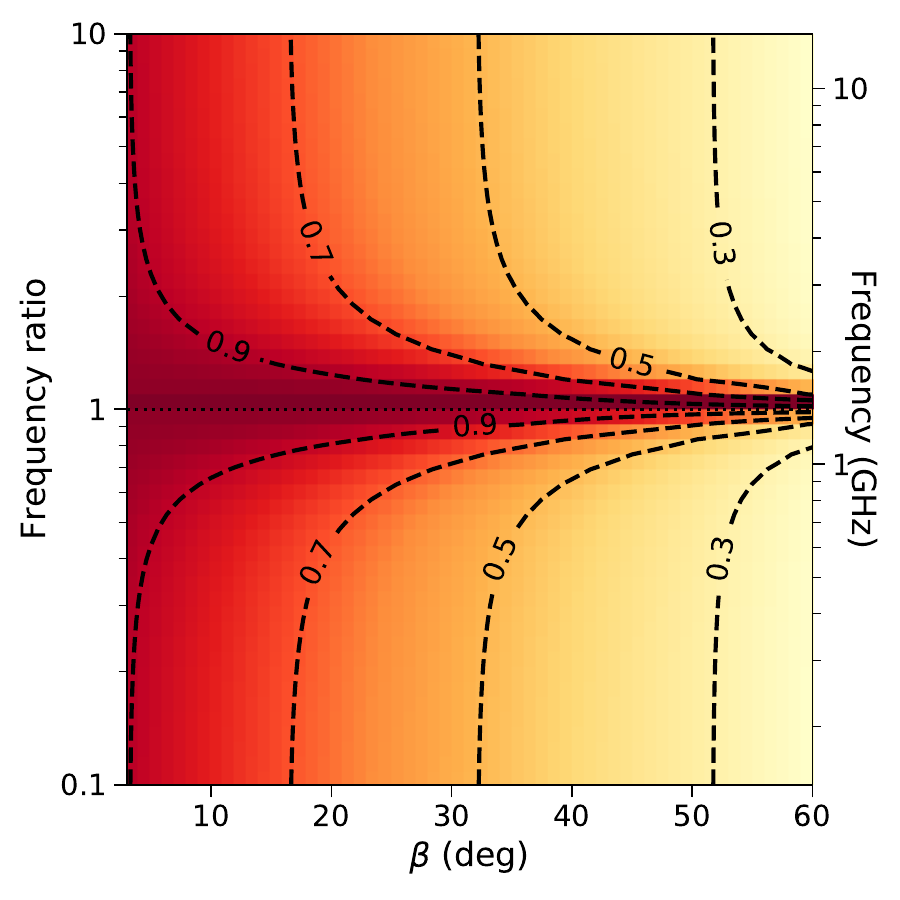}

    \vspace{-0.1cm}
    \caption{Heat map showing the measurement uncertainties of $\varpi$ from simulated data consisting of two frequency bands. The simulations were conducted for different ecliptic latitudes and ratios of the frequency of the two bands. The errors were normalized by that at frequency ratio of 1 and with the same $\beta$ angle. Here we assumed $\varpi=1$\,mas and $n_{\rm sw}$=8\,cm$^{-3}$. The dotted horizontal line in the lower panel represents frequency ratio of unity and thus normalized uncertainty of 1. \label{fig:PXerr_multif}}
\end{figure}

The above studies show that for favourable timing precision and ecliptic latitude of the pulsar, the astrometric and solar-wind parameters can be measured jointly even if the data are collected at a single frequency. This means that for dataset with poor multi-frequency coverage (which could be the case for legacy pulsar timing dataset), fitting for a constant solar-wind density could still be attempted so as to avoid the bias in the estimate of astrometric parameters. Of course, collecting data from multi-frequency observations (usually the case in modern pulsar timing experiments) could mitigate the correlation between $\varpi$ and $n_{\rm sw}$ and improve their measurement precision. For a detailed investigation along this line, we simulated datasets each containing two frequency bands. In each iteration, one dataset were generated centred at 1.4\,GHz and the other at an alternative frequency (which varies between iterations). Figure~\ref{fig:PXerr_multif} shows the measurement uncertainties of $\varpi$ from a timing fit to the data for various ecliptic latitudes of the pulsar and the frequency ratios of the two bands. The results demonstrate a clear increase in the measurement precision of $\varpi$ as the frequency separation between the two bands becomes larger, compared to the case where the same amount of observations are all spent at a single frequency. The extent of the improvement depends on the ecliptic latitude of the pulsar. For lower ecliptic latitude, e.g., $\beta<10$\,deg, the precision improves marginally with the extension of frequency coverage, up to a level of approximately 20\%. The improvement is more prominent for higher ecliptic latitude cases. For instance, at $\beta$ of 30\,deg, the decrease in uncertainty can exceed 40\%, while at $\beta$ of 60\,deg, it can be as much as 60\% when the frequency separation of the two bands is only 20\%. This demonstrates that conducting multi-frequency observations is more important for pulsars at high ecliptic latitudes, so as to yield a better precision in parallax measurement.

\begin{figure}
\centering
	\includegraphics[scale=0.7]{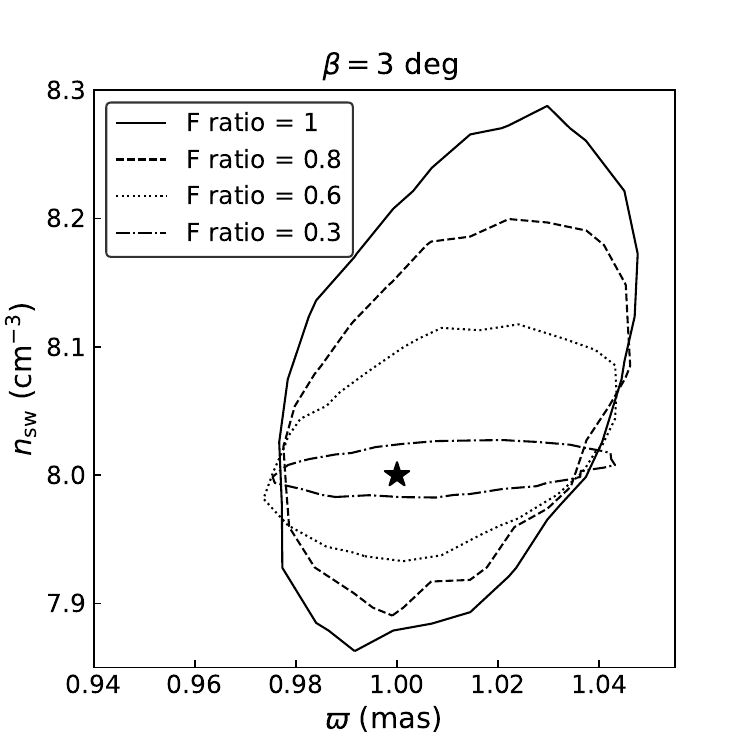}

 
   \includegraphics[scale=0.7]{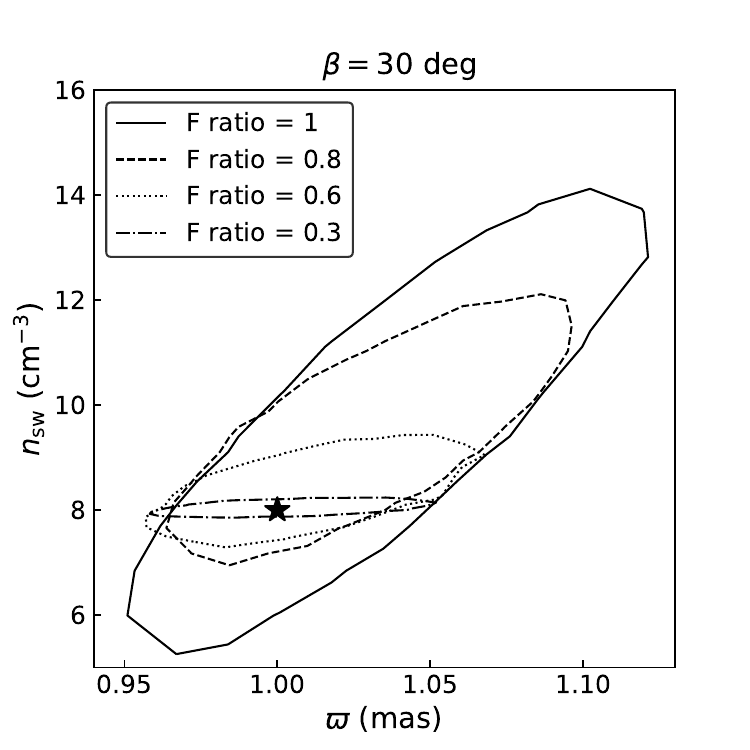}

    \vspace{-0.1cm}
    \caption{Comparison of joint posterior 1-$\sigma$ contours of $\varpi$ and $n_{\rm sw}$ for datasets with different frequency ratio of the two frequency bands. The upper and lower panels show the results for ecliptic latitude of $3$\,deg and $30$\,deg, respectively. The stars represents the injected value of the two parameters, $\varpi=1$\,mas and $n_{\rm sw}=8$. \label{fig:PXNE_contour}}
\end{figure}

Figure~\ref{fig:PXNE_contour} illustrates the improvement in constraining $\varpi$ and $n_{\rm sw}$ as the frequency coverage is increased, by showing the joint posterior 1-$\sigma$ contours of $\varpi$ and $n_{\rm sw}$ from the Bayesian timing analysis. It can be seen that the contours become tighter and more orthogonal as the frequency ratio of the two bands deviates from unity. This is a result of the decrease in the correlation between these two parameters, as expected from Figure~\ref{fig:fcov_rms_cc}. The reduced correlation allows for more precise and independent measurements of $\varpi$ and $n_{\rm sw}$, highlighting the benefits of increased frequency separation in multi-frequency observations. It is also notable that the improvement on the parameter constraint is more significant for $\beta=$30\,deg than $\beta=$3\,deg. This echos the conclusion above that multi-frequency observations is crucial for precisely measuring $\varpi$ and $n_{\rm sw}$, if the pulsar is at a high ecliptic latitude.

\begin{figure}
\centering
    \includegraphics[scale=0.55]{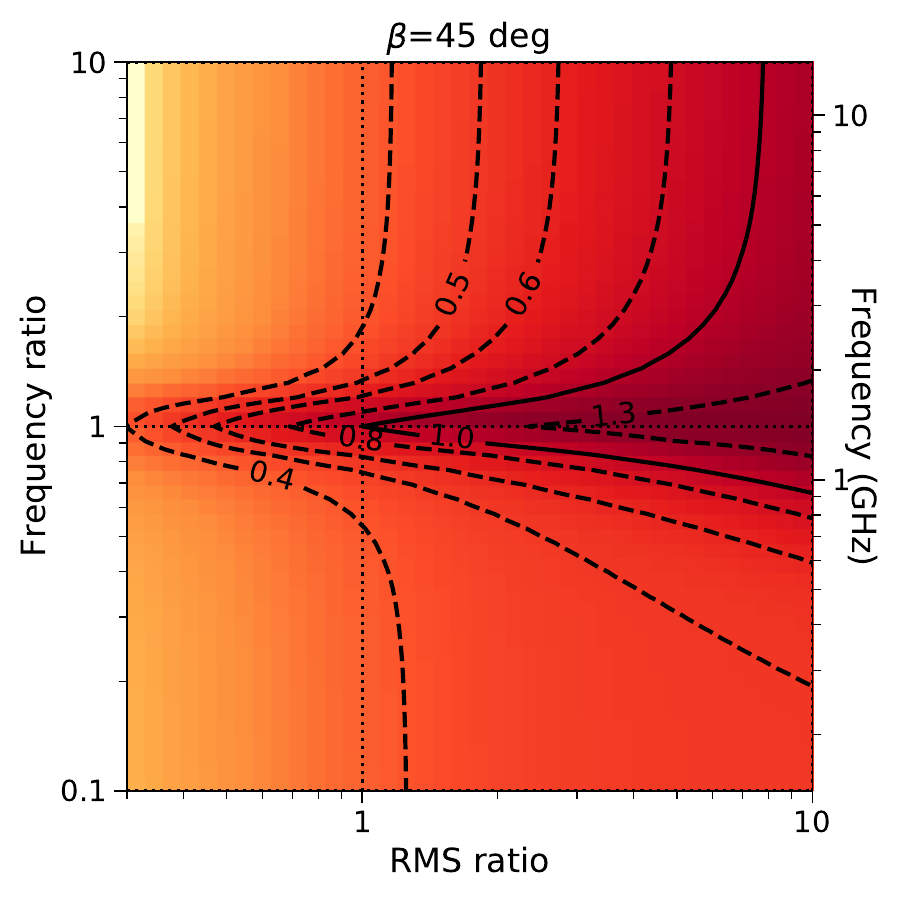}
    \vspace{-0.1cm}
    \caption{Heat map showing the normalised measurement uncertainties of $\varpi$ from simulated data consisting of two frequency bands, with different frequencies and timing precision. The uncertainties of $\varpi$ were relative to that at frequency and rms ratio of unity. Here we assumed $\varpi=1$\,mas and $n_{\rm sw}$=8\,cm$^{-3}$, $\beta=45$\,deg. The dotted horizontal and vertical lines represents frequency and rms ratio of unity, respectively. \label{fig:err_F_rms_beta45}}
\end{figure}

The above simulations assume equal timing precision at the two frequency bands, while in reality it is likely to be different. Figure~\ref{fig:err_F_rms_beta45} shows the results of the simulation where the alternative band has a different rms residual. Clearly, if the timing precision at the alternative band is higher, there is always improvement in the measurement precision of the parallax, compared to the case of having the same frequency for the two bands. However, if the timing precision of the alternative band is lower, for pulsars at a moderate ecliptic latitude (here $\beta=45$\,deg) or higher, the parallax measurement could still be significantly improved. Take the reference frequency of 1.4\,GHz as an example from the figure. For a higher frequency of 2.5\,GHz, the uncertainty in $\varpi$ is still lower even if the timing precision is a factor of a few worse. For frequencies lower than 900\,MHz, the precision in $\varpi$ is higher for rms up to an order of magnitude higher. 

\subsection{A note on observations of pulsars at a very low ecliptic latitude}

\begin{figure}
\centering
	\includegraphics[scale=0.6]{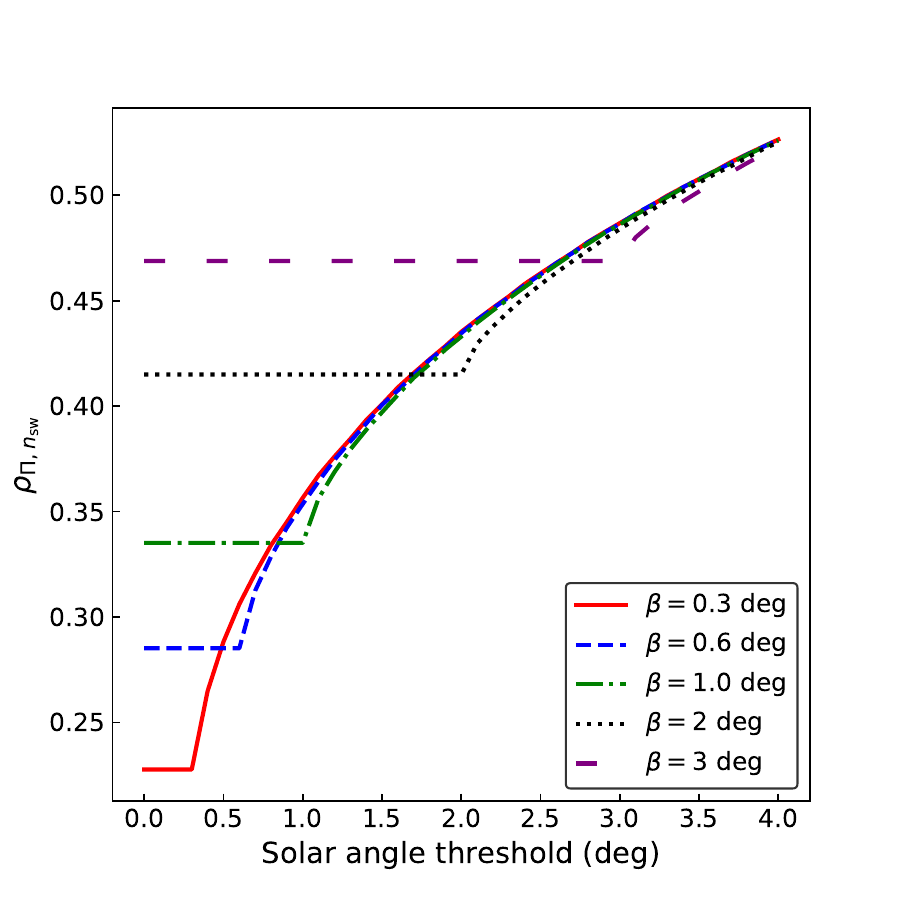}
    \caption{Correlation coefficient of $\varpi$ and $n_{\rm sw}$ as a function of telescope solar angle constraint, for sources at various ecliptic latitudes.  
    \label{fig:solar_angle_limit}}
\end{figure}

The analysis above assumes a complete sampling of the solar-wind delay feature around the solar conjunction, where the delay signal is the most prominent. This is a realistic assumption except for pulsars at very low ecliptic latitudes, where practical limitations arise due to the minimum possible solar angle of the source during observations. Firstly, the angular size of the Sun from Earth is 0.524--0.542\,deg. This means that for pulsars at $\beta<0.3$\,deg, they cannot be observed for a short period of time ($\sim$2 days) across the solar conjunction because they will be behind the Sun. Additionally, reflecting radio telescopes often have a limit on how close they can point to the Sun to avoid overheating in the receiver focal cabin due to the concentration of optical and infrared radiation\footnote{For instance, the limit is approximately 3\,deg for observations at 1.4\,GHz with the Effelsberg Radio Telescope.}. This constraint means data cannot be collected when the pulsar line-of-sight is too close to the Sun. Some telescopes that are only able to reflect and concentrate long-wavelength electromagnetic signals in the radio spectrum but not optical and infrared light, would be able to observe closer to the Sun. However, the limiting factor in this case is that the radio signal from the Sun may be included in the side-lobe of the beam, significantly contributing to system temperature and reducing observation sensitivity. For instance, the LOw Frequency ARray (LOFAR) is able to collect useful data for angular separation of the pulsar to the Sun down to approximately 1\,deg \citep{tsb+21}. 

The lack of data samples around solar conjunction likely impacts the correlation between astrometric and solar-wind parameters and thus affects their measurement precision during practical data analysis. Figure~\ref{fig:solar_angle_limit} shows the correlation coefficient of $\varpi$ and $n_{\rm sw}$, obtained from data simulated with a given solar-angle limit and at various small ecliptic latitudes. Clearly, in all cases the correlation begins to increase when the solar-angle limit is larger than the ecliptic latitude of the pulsar, resulting in incomplete sampling of the solar-wind delay feature. This leads to less precise measurements of both astrometric and solar-wind parameters compared to the idealistic scenario. It is also worth noticing, that for a certain solar-angle limit, $\rho_{\varpi,n_{\rm sw}}$ is approximately the same for all ecliptic latitudes smaller than this limit. This is likely because the solar-wind delay waveforms in these case are all very similar, once features around the solar conjunction are excluded. 

\section{Fitting for solar-wind density with the EPTA DR2 dataset} \label{sec:real}

\begin{figure*}
\centering
	\includegraphics[scale=0.38]{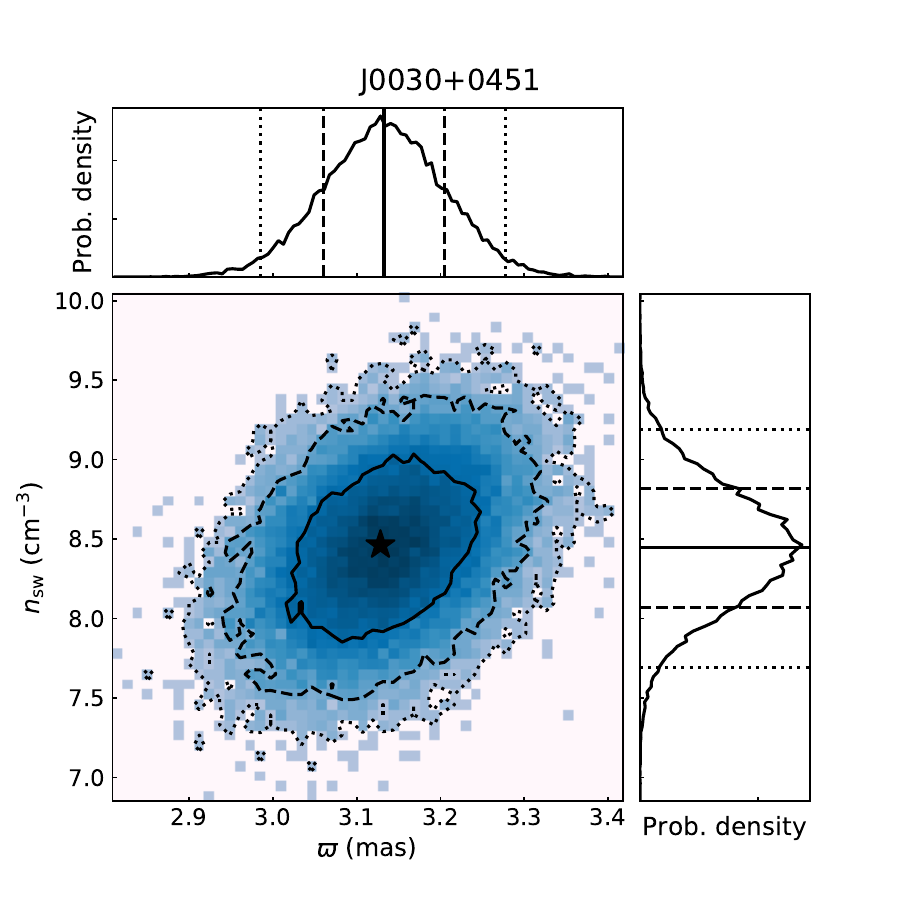}
    \includegraphics[scale=0.38]{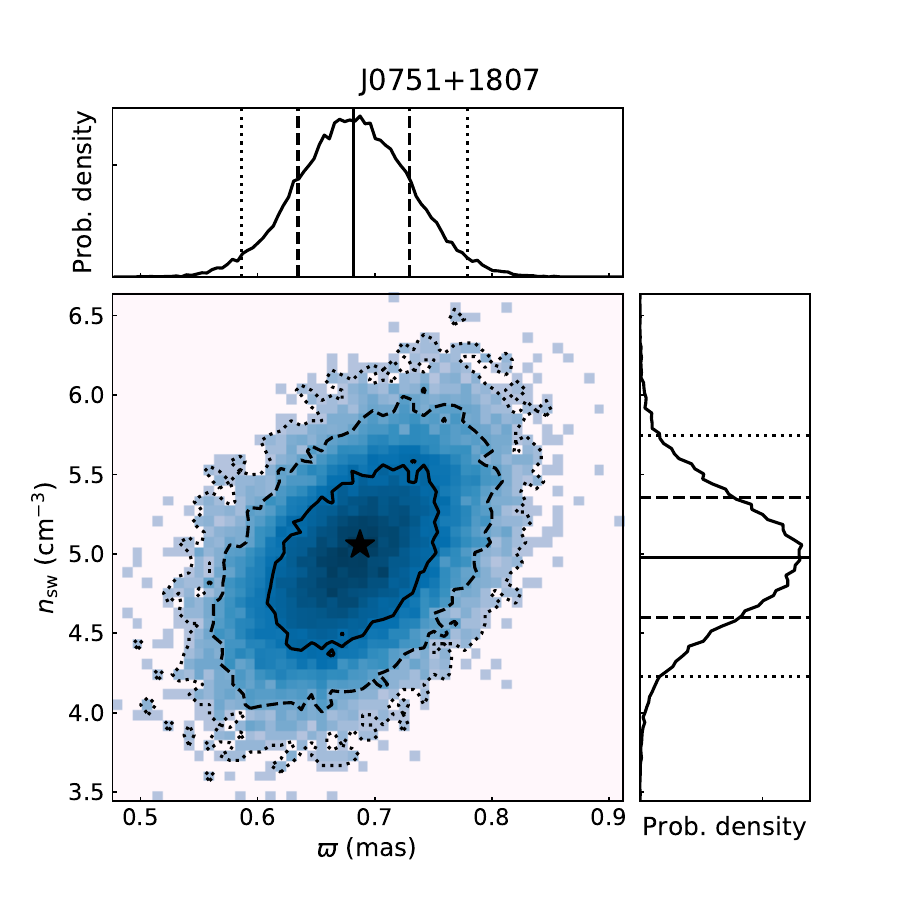}
    \includegraphics[scale=0.38]{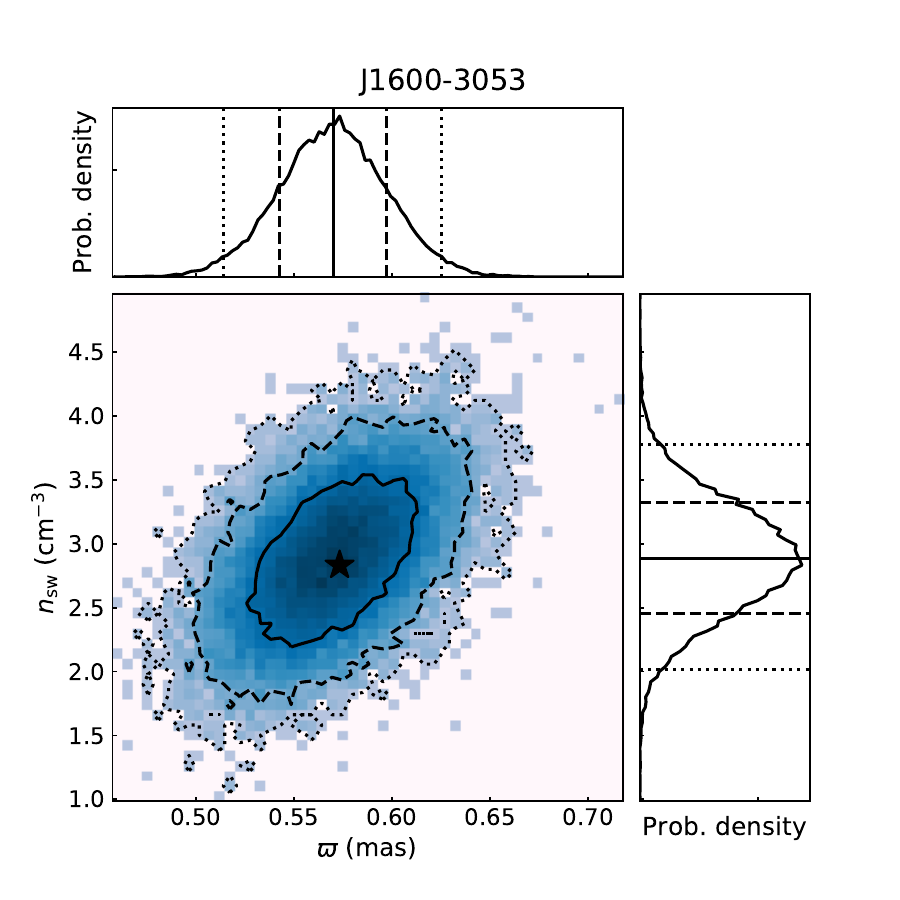}
    \includegraphics[scale=0.38]{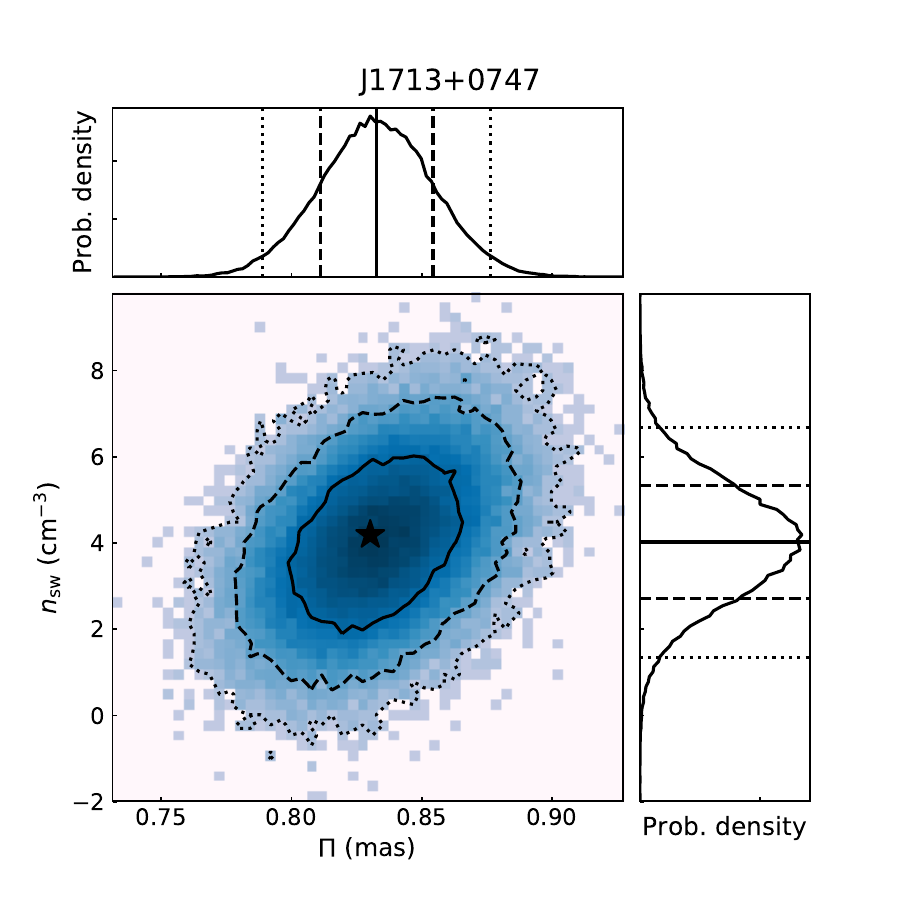}
    \includegraphics[scale=0.38]{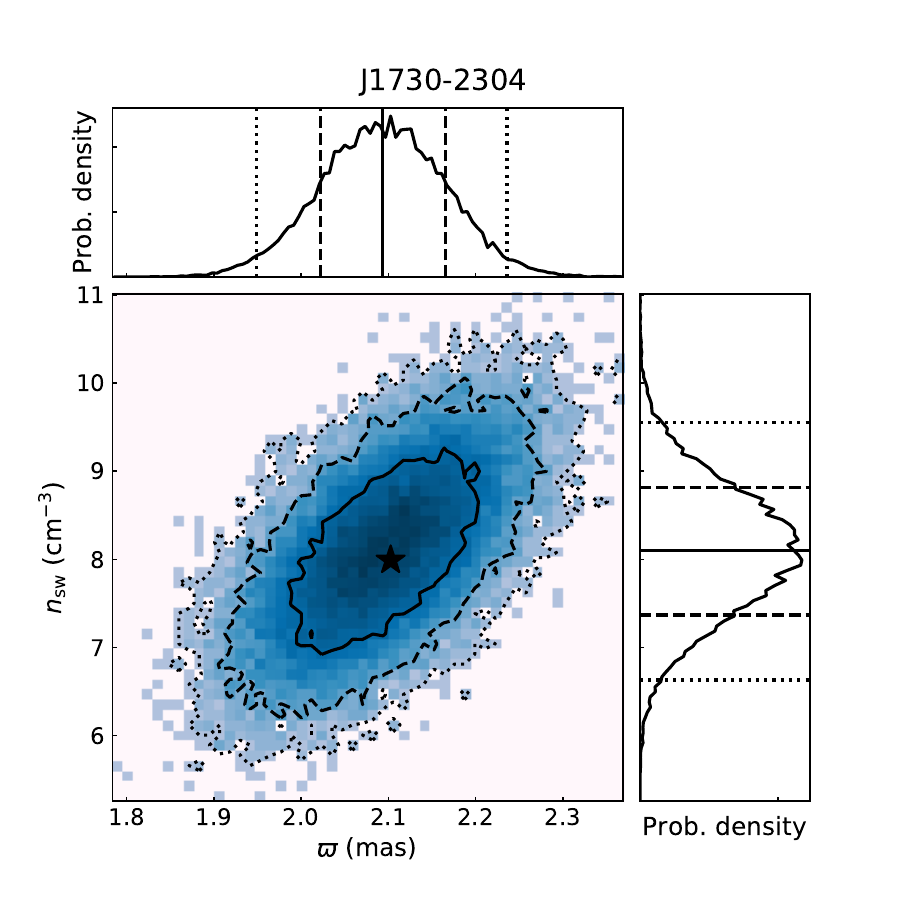}
    \includegraphics[scale=0.38]{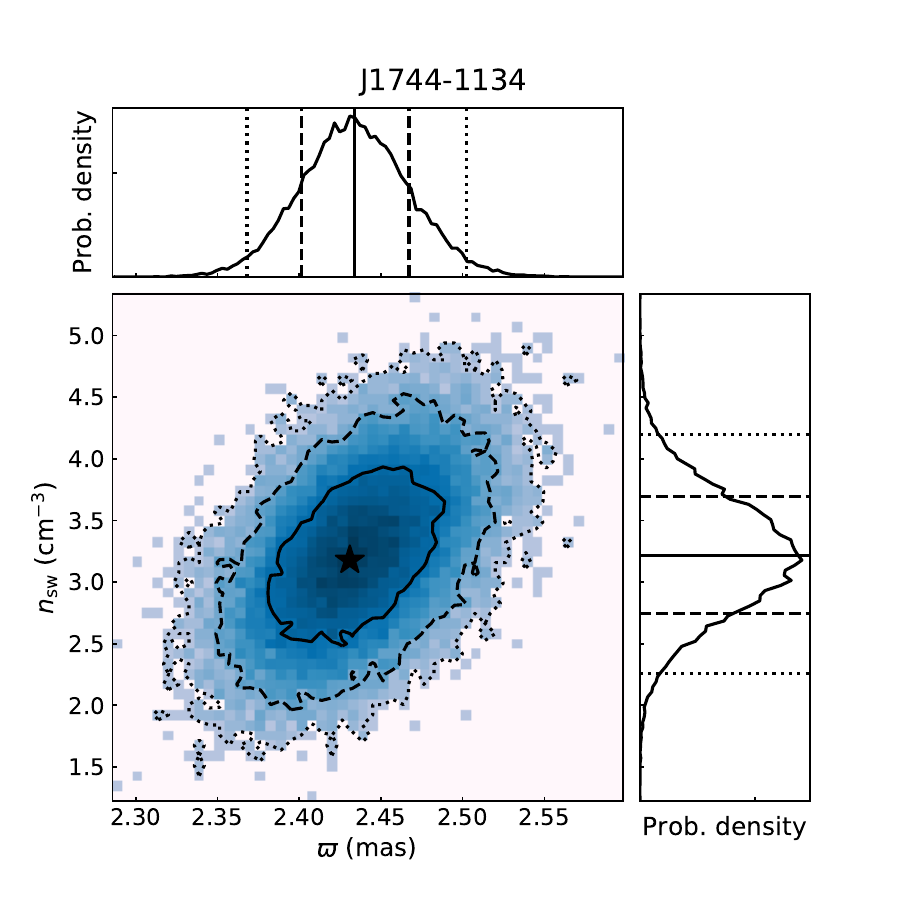}
    \includegraphics[scale=0.38]{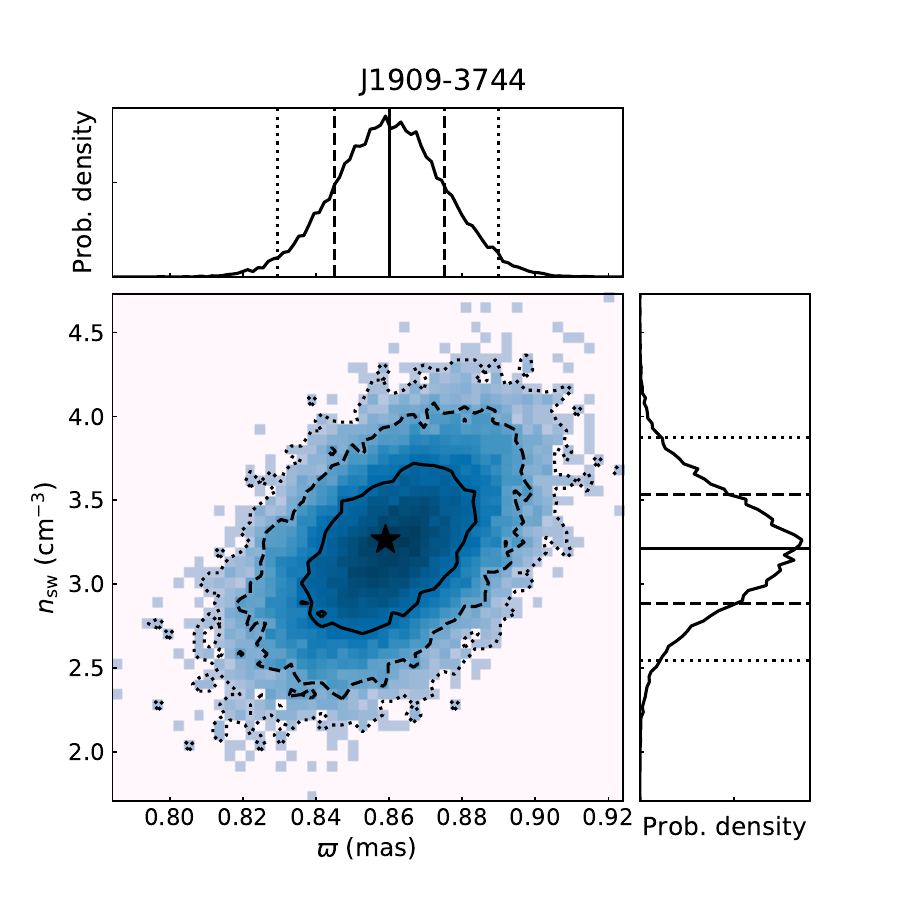}
    \caption{Joint posterior distributions of $\varpi$ and $n_{\rm sw}$ measured from EPTA `DR2full` dataset. The stars in the panels correspond to the maximum likelihood value of the parameters. \label{fig:PXNE_DR2}}
\end{figure*}

Following the findings from simulation studies presented above, in this section we reprocessed data from the second data release of the European Pulsar Timing Array (EPTA), to incorporate modelling of $n_{\rm sw}$ in the timing analysis. We used the version `DR2full` dataset \citep[see Appendix A of][]{EPTA+2023a} that contains data from both the legacy and new-generation pulsar backends in the EPTA, mostly at gigahertz frequencies. The analysis focused on a selection of pulsars where $\varpi$ was measured and the anticipated scale of the solar-wind delay is comparable to the achieved timing precision, allowing for reasonable constraints on $n_{\rm sw}$. The deterministic timing model described in \cite{EPTA+2023a} were employed for the timing analysis. To account for the red noise presented in the data, we used the customized noise model determined by \cite{EPTA+2023b} for each pulsar. In short, three types of red-noise model were considered to be included: the chromatic red noise caused by DM and scattering variation, respectively, and the achromatic red-noise component. In all models, the red signal was assumed to be a stationary, stochastic process with a power-law spectrum in the form of
\begin{equation}
S(\mathscr{f})=\frac{A^2}{C_{\rm 0}}\left(\frac{\mathscr{f}}{\mathscr{f}_{\rm r}}\right)^{-\gamma},
\label{eq:rednoise}
\end{equation}
where $S(\mathscr{f})$, $A$, $\gamma$, $\mathscr{f}_{\rm r}$ are the power spectral density as a function of Fourier frequency $\mathscr{f}$, spectral amplitude, spectral index and the reference Fourier frequency (set to 1\,yr$^{-1}$), respectively. Here, the spectral amplitude relates to the observing radio frequency as $A\propto\nu^{-\alpha}$, where the chromatic index $\alpha$ is $2,4,0$ for the case of DM variation, scattering variation and achromatic noise, respectively. The constant $C_{\rm 0}$ equals to $12\pi^2$ for scattering variation and achromatic red noise, and is $k^2_{\rm DM}$ for DM variation where $k_{\rm DM}=2.41\times10^{-4}$\,cm$^{-3}$\,pc\,MHz$^2$\,s$^{-1}$ is the DM constant. Again, the \textsc{temponest} software package was utilized to perform the Bayesian timing analysis and generate the posterior distributions of the model parameters \citep{lah+14}. Both $\varpi$ and $n_{\rm sw}$ were sampled simultaneously with the other timing and noise model parameters. The results of parameter measurements are summarized in Table~\ref{tab:dr2_ne_sw}, with the joint posterior distributions of $\varpi$ and $n_{\rm sw}$ presented in Figure~\ref{fig:PXNE_DR2}. 

In the analysis of \cite{EPTA+2023a}, $n_{\rm sw}$ was fixed to 7.9\,cm$^{-3}$ for all pulsars \citep{mca+19}, except for PSR~J1022+1001 where a fitted value of $11.1\pm0.3$\,cm$^{-3}$ was obtained. For PSRs J0030+0451, J0751+1807, J1730$-$2304, our timing analysis yields values very close to the previously fixed $n_{\rm sw}$. The measured positions and parallaxes are highly consistent with those from \cite{EPTA+2023a}, though of slightly lower precision. This outcome is entirely expected due to the inclusion of $n_{\rm sw}$ in the timing fit and the resulting correlation between $\varpi$, $n_{\rm sw}$. 

Meanwhile, for PSRs J1600$-$3053, J1713+0747, J1909$-$3744 and J1744$-$1134, the fitted $\varpi$ and $\beta$ values from our analysis are smaller than those obtained in \cite{EPTA+2023a}, and their fitted $n_{\rm sw}$ values are also correspondingly less than 7.9\,cm$^{-3}$ as anticipated from our simulations. In fact, the inclusion of $n_{\rm sw}$ in the timing fit delivers parallax measurements more consistent with those reported elsewhere, such as in \cite{NG15+a} where the `DMX` model was employed to capture chromatic effects in the data including both DM and solar wind, and \cite{rsc+21} where a fixed $n_{\rm sw}$ of 4\,cm$^{-3}$ was used in the timing analysis. For instance, in the case of PSR~J1713+0747, the fitted value of $n_{\rm sw}$ is only marginally constrained at $4.0\pm1.3$\,cm$^{-3}$. At the same time, the fitted $\varpi$ value now decreases from $0.88\pm0.01$ to $0.83\pm0.01$\,mas, which is highly consistent with $0.833\pm0.015$\,mas in \cite{NG15+a} and closer to $0.763\pm0.021$\,mas in \citep{rsc+21}. The decrease of 0.05\,mas generally matches the prediction in Figure~\ref{fig:PXbias} for a -3.9\,cm$^{-3}$ offset in $n_{\rm sw}$. In all cases of Table~\ref{tab:dr2_ne_sw}, the fitted values of ecliptic longitude from our analysis are the same as those obtained in \cite{EPTA+2023a}.

The relation between our obtained $n_{\rm sw}$ values to the ecliptic latitude of the pulsar matches the expectations from \citet{sct+24}. For pulsars at very low ecliptic latitudes ($\beta\lesssim5$\,deg: PSRs J0030+0451, J0751+1807, J1730$-$2304) the solar-wind density is higher because the slow (and denser) solar wind is the dominant component along the line-of-sight (LOS). For those at higher ecliptic longitudes (here PSRs J1600$-$3053, J1713+0747, J1744$-$1134 and J1909$-$3744), the value is lower since the LOS could pass through either primarily the fast solar wind (around solar minima) or a mixture of the two components (around solar maxima). For PSRs J0030+0451, J1730$-$2304 and J1744$-$1134, our measured values from EPTA `DR2full` dataset show good consistency with those reported by \citet{sct+24} using low-frequency ($\sim200$\,MHz) observations with LOFAR. Nonetheless, the interpretation of these measurements should be taken with caution as pointed out later. 

The noise-model parameters measured from our analysis are highly consistent with those obtained in \cite{EPTA+2023b}. Notably, the achromatic red-noise measurements remain unchanged after the inclusion of $n_{\rm sw}$ in the timing fit. Therefore, we do not expect significant impact on the gravitational-wave search results presented in \cite{EPTA+2023c} by fixing the $n_{\rm sw}$ value. 

\begin{table*}
	\centering
	\caption{Measurement of timing and noise parameters for a selection of pulsars from our analysis and \citet{EPTA+2023a} (with subscript `DR2'), with solar-wind density as an deterministic timing parameter. From the top to bottom, the raws are ecliptic latitude ($\beta$, $\beta_{\rm DR2}$), parallax ($\varpi$, $\varpi_{\rm DR2}$), solar-wind density ($n_{\rm sw}$), amplitude and spectral index of DM variation ($A_{\rm DM}$, $\gamma_{\rm DM}$), achromatic red noise ($A_{\rm red}$, $\gamma_{\rm red}$) and scattering variation ($A_{\rm Scat}$, $\gamma_{\rm Scat}$). The parallax measurements from \citet{EPTA+2023a} ($\varpi_{\rm DR2}$) were also presented as a comparison. The values in the table represent the median of the posterior distribution from the Bayesian timing analysis with 1$-\sigma$ credible interval.}
	\label{tab:dr2_ne_sw}
	\begin{tabular}{lccccccc}
		\hline
  \hline
		PSR Jname & J0030+0451 & J0751+1807 & J1600$-$3053 & J1713+0747 & J1730$-$2304 & J1744$-$1134 & J1909$-$3744 \\
  \hline
$\beta$ (deg) & $1.4457(14)$ & -2.8075798(2) & -10.07183448(3) & 30.700361492(5) & 0.189(14) & 11.80520318(3) & -15.15550397(1) \\
$\beta_{\rm DR2}$ (deg) & $1.4457(14)$ & -2.8075795(2) & -10.07183439(4) & 30.700361485(5) & 0.189(14) & 11.80520311(3) & -15.15550393(1) \\
$\varpi$ (mas)  &$3.13(7)$  & $0.68(5)$ & $0.57(3)$ & $0.83(2)$ & $2.09(7)$ & $2.43(3)$ & $0.86(2)$ \\
$\varpi_{\rm DR2}$ (mas) &$3.09(6)$ & $0.85(4)$ & $0.72(2)$ & $0.88(1)$ & $2.08(6)$ & $2.58(3)$ & $0.94(2)$\\
$n_{\rm sw}$ & $8.4(4)$ & $5.0(4)$ & $2.9(4)$ & $4.0(1.3)$ & $8.1(7)$ & $3.2(5)$ & $3.2(3)$ \\
\hline
$A_{\rm DM}$ & -- & $-11.8_{-0.1}^{+0.1}$ & $-12.2_{-0.2}^{+0.2}$ & $-11.88^{+0.04}_{-0.04}$ & $-11.6_{-0.3}^{+0.3}$ & $-11.84_{-0.04}^{+0.04}$ & $-12.12_{-0.04}^{+0.04}$ \\ \noalign {\smallskip}
$\gamma_{\rm DM}$ & -- & $3.0_{-0.3}^{+0.3}$ & $4.1_{-0.5}^{+0.5}$ & $1.67^{+0.02}_{-0.02}$  & $2.2_{-0.8}^{+0.8}$  & $1.34_{-0.16}^{+0.16}$ & $1.85_{-0.15}^{+0.16}$ \\ \noalign {\smallskip}
$A_{\rm red}$ & $-14.9_{-0.6}^{+0.5}$  & -- & -- & $-14.2^{+0.2}_{-0.2}$ & -- & $-15.4_{-1.1}^{+1.2}$ & $-15.0_{-0.5}^{+0.4}$ \\ \noalign {\smallskip}
$\gamma_{\rm red}$  & $5.42_{-0.87}^{+1.00}$ & -- & -- & $3.24^{+0.57}_{-0.52}$ & -- & $5.4_{-1.8}^{+1.7}$ & $5.2_{-0.9}^{+1.1}$ \\ \noalign {\smallskip}
$A_{\rm Scat}$ & -- & -- &$-13.36_{-0.04}^{+0.04}$ & -- & -- & -- & -- \\ \noalign {\smallskip}
$\gamma_{\rm Scat}$ & -- & -- &$1.40_{-0.14}^{+0.14}$ & -- & --  & -- & --\\
\hline 
\hline
	\end{tabular}
\end{table*}


\section{Conclusions and discussions} \label{sec:dis}

In this paper, we investigated the impact on measurement of astrometry in pulsar timing by the solar wind effect, which could pose a potential issue for precisely measuring pulsar astrometry and distance. Using both theoretical calculation and mock-data simulations, we demonstrated a significant correlation between the pulsar position and parallax parameters, and the solar-wind density parameter, $n_{\rm sw}$. This correlation strongly depends on the ecliptic latitude of the pulsar. We showed that using a fixed $n_{\rm sw}$ with an arbitrary value in the timing analysis could introduce significant bias in the estimated pulsar position and parallax, leading to inaccuracies in the pulsar distance measurement. The significance of the bias strongly depends on the ecliptic latitude of the pulsar and the timing precision of the dataset. For pulsars with favourable timing precision and ecliptic latitude, the astrometric and solar-wind parameters can be measured jointly with other timing parameters already using single-frequency data, which will avoid the bias in the estimate of astrometric parameters. The correlation between astrometric and solar-wind parameters can be mitigated by using data with multi-frequency coverage, which also greatly improves the measurement precision of these parameters. This is in particular significant for pulsars at a medium or high ecliptic latitude ($\beta\gtrsim30$\,deg). Furthermore, we reprocessed EPTA `DR2full` dataset to include the solar-wind effect in the timing fit for a selection of pulsars, which resulted in significant measurements of both parallax and solar-wind density. For pulsars where $n_{\rm sw}$ differed from the fixed value of 7.9\,cm$^{-3}$ in the EPTA DR2 analysis, the parallax measurement became more consistent with values reported elsewhere. For sources overlapping with the analysis in \citet{sct+24} using low-frequency observations, our $n_{\rm sw}$ measurements align well with theirs.

As pointed out by various previous work, the spherical model in Eq.~\ref{eq:sol_delay} is not complete in describing the solar-wind effect. Most importantly, it neglects the time variation in the solar-wind density which can be caused by both the long-term (11-year) solar cycles and short-term stochastic process in the fast solar wind component \citep{stz20,tsb+21}. For instance, recently \citet{sct+24} observed complex time variability in the solar-wind intensity for several PTA pulsars using data collected with LOFAR. Therefore, our measurements using the model in Eq.~(\ref{eq:sol_delay}) should be seen as representing a time average of the solar-wind density over many years. Caution needs to be taken when using these measurements to derive other physical properties of the solar wind. Nonetheless, it was shown that the EPTA DR2 data, mostly at frequencies higher than 1\,GHz, is likely not to be sensitive to such time variation \citep{nkt+24}. The model should also serve the primary purpose of our simulation studies to understand the correlation between astrometric and solar-wind parameters. It is anticipated that being able to model the solar-wind time variability could mitigate the degeneracy between astrometric and solar-wind parameters. 


Our simulations have shown that the measurability of $\varpi$ and $n_{\rm sw}$ strongly depends on several variables, including the pulsar's ecliptic latitude, observing frequency, timing precision of the data, etc.. At 1.4\,GHz, the typical frequency of most precision timing observations, for a pulsar at a very low ecliptic latitude of $\beta\lesssim15$\,deg with a timing precision of the order 1\,$\upmu$s, it should always be recommended to include solar-wind modelling in the timing analysis to avoid significant bias in the fitted astrometric parameters. This also applies to pulsars at higher ecliptic latitude of 15\,deg$\lesssim\beta\lesssim30$\,deg, if the timing precision is of the order 100\,ns. For pulsars with $\beta\gtrsim30$\,deg, it is possible at this frequency to encounter a situation where $\varpi$ is well measured with the data but $n_{\rm sw}$ may not. In this scenario, while fitting for $n_{\rm sw}$ still remains a conservative option, using a given $n_{\rm sw}$ may also be feasible, which may introduce a bias in the astrometric parameters that is tolerable within the measurement uncertainty. If this approach is employed in the timing analysis, based on the results from \citet{sct+24}, a fixed value (e.g., 4\,cm$^{-3}$) lower than the gross average of many PTA pulsars (7.9\,cm$^{-3}$) obtained in \cite{mca+19}, would likely provide a more realistic representation of the actual $n_{\rm sw}$ for pulsars falling in this ecliptic latitude range.

Our work shows complexity in obtaining robust astrometry measurements in pulsar timing analysis, due to the significant correlation between the astrometric and solar-wind parameters. In addition, \cite{mcc13} has also demonstrated that timing measurement of astrometry can be contaminated by red noise in the timing data, if it is not properly accounted for. Therefore, to obtain unbiased astrometric measurements with pulsar timing, it would be recommended to conduct both the timing and noise analysis at the same time. Meanwhile, measurements from independent experiments, such as the VLBI observations \citep[e.g.,][]{dds+23} or the scintillation studies \citep[e.g.,][]{rch+19}, could be used to verify the timing results and even help to break the degeneracy among the model parameters. 

In line with previous studies \citep[e.g.,][]{lbj+12,lmc+18,dvt+20}, our results highlight the importance of achieving high timing precision at higher radio frequencies in pulsar timing analysis to correct for chromatic effects, such as those caused by the solar wind. However, this precision is not easily achieved and requires dedicated efforts. Since pulsars are typically steep-spectrum radio sources \citep[e.g.,][]{jvk+18}, their flux density is significantly higher at lower frequencies, making it easier to achieve sufficient timing precision for modeling chromatic effects without placing a strong demand on system sensitivity. In contrast, at higher radio frequencies, the sharp drop in flux density necessitates the use of the largest radio telescopes or extended integration times to achieve comparative timing precision. Despite these challenges, using the most sensitive telescopes for high-frequency timing observations aligns with recommendations in other studies. Notably, timing precision at canonical timing frequencies (e.g., 1.4\,GHz) for many MSPs observed with the largest radio telescopes is often limited by intrinsic white noise caused by the instability of pulsar emission on short timescales, known as ``jitter noise'' \citep[e.g.,][]{lvk+11,pbs+21,www+24}. In such cases, conducting timing observations at higher frequencies is more beneficial, as it allows for a precision that less sensitive telescopes cannot achieve. Therefore, using more sensitive telescopes at higher radio frequencies is advantageous for multiple reasons.

Finally, our simulation studies have focused on assessing the impact of the solar-wind effect on astrometric measurements in precision pulsar timing, particularly under the assumption of canonical timing observations around gigahertz radio frequencies with weekly cadence. We emphasise the significant value of ongoing low-frequency ($<800$\,MHz) pulsar timing programs, such as those conducted with the Canadian Hydrogen Intensity Mapping Experiment (CHIME) Telescope \citep{CHIME+21} and LOFAR \citep{tvs+19}, for studying chromatic effects in timing analysis such as the solar wind. Meanwhile, due to limitations in achievable timing precision and baseline, these low-frequency data alone may not suffice for measuring pulsar astrometry or achieving competitive precision. Therefore, combining low-frequency and canonical pulsar timing data would provide the optimal solution for simultaneously and robustly measuring both pulsar astrometry and the solar-wind properties.

\section*{Acknowledgements}
We thank Joris Verbiest and Paulo Freire for reading the manuscript and providing constructive comments. KL and MK acknowledge funding by the MPG-CAS LEGACY programme. JA acknowledges support from the European Commission (ARGOS-CDS; Grant Agreement number: 101094354). A.C. acknowledges financial support provided under the European Union's Horizon Europe ERC Starting Grant "A Gamma-ray Infrastructure to Advance Gravitational Wave Astrophysics" (GIGA, Grant Agreement: 101116134). The Effelsberg 100-m telescope is operated by the Max-Planck-Institut f{\"u}r Radioastronomie. The Nan\c{c}ay radio Observatory is operated by the Paris Observatory, associated with the French Centre National de la Recherche Scientifique (CNRS), and partially supported by the Region Centre in France. We acknowledge financial support from ``Programme National de Cosmologie and Galaxies'' (PNCG), and ``Programme National Hautes Energies'' (PNHE) funded by CNRS/INSU-IN2P3-INP, CEA and CNES, France. We acknowledge financial support from Agence Nationale de la Recherche (ANR-18-CE31-0015), France. The Westerbork Synthesis Radio Telescope is operated by the Netherlands Institute for Radioastronomy (ASTRON) with support from the Netherlands Foundation for Scientific Research (NWO). Pulsar research at the Jodrell Bank Centre for Astrophysics and the observations using the Lovell Telescope are supported by a Consolidated Grant (ST/T000414/1) from the UK's Science and Technology Facilities Council (STFC).

\section*{Data Availability}
 
The EPTA `DR2full` dataset are available at: \url{https://zenodo.org/records/8300645}.



\bibliographystyle{mnras}
\bibliography{psrrefs,modrefs,crossrefs,2024} 




\appendix

\section{Evaluation of the parameter covariance} \label{sec:app}
In pulsar timing analysis, the time transformation formula, i.e., the timing model, consists of a sequence of time delay terms that transfers the TOAs of pulsar signals at the observatory to those in the pulsar initial frame, as given by Eq.~(\ref{eq:timing_model}). The correlation between model parameters can be evaluated by calculating the covariance matrix in the least-square fit by following the method presented in \cite{bt76} and \cite{dd86}. Here for simplification, when estimating the correlation between $\varpi$ and  $n_{\rm sw}$, we only consider the solar-system Roemer delay term and the two terms introduced by annual parallax and solar wind, respectively. These terms include all timing parameters that are expected to exhibit strong correlation with $\varpi$ and $n_{\rm sw}$. Using ecliptic coordinate, namely $\beta$, $\lambda$, to describe the sky position of the pulsar (and neglecting proper motion), the derivative timing formula can be written as \citep{bt76}:
\begin{equation}
    \mathcal{R} = -\delta N_{\rm 0} / \nu + P \cdot \delta\varpi + Q \cdot \delta n_{\rm sw} + R \cdot \delta\beta + S \cdot \delta \lambda
\end{equation}
where $P$, $Q$ are given by Eq.~(\ref{eq:PQ}), and $R$, $S$ are written as:
\begin{eqnarray}
R&=&\frac{\partial\Delta_{\rm R\odot}}{\partial\beta} + \frac{\partial\Delta_{\rm p}}{\partial\beta} + \frac{\partial\Delta_{\rm sol}}{\partial\beta} \\
S&=&\frac{\partial\Delta_{\rm R\odot}}{\partial\lambda} + \frac{\partial\Delta_{\rm p}}{\partial\lambda} + \frac{\partial\Delta_{\rm sol}}{\partial\lambda}.
\end{eqnarray}
The derivative terms are expressed by
\begin{eqnarray}
\frac{\partial\Delta_{\rm R\odot}}{\partial\beta} & = & \frac{|\Vec{r}(E)|\cos{(A_{\rm T}-\lambda)}\sin{\beta}}{c} \\
\frac{\partial\Delta_{\rm p}}{\partial\beta} & = & \frac{\varpi|\Vec{r}(E)|^2}{2c}\cos^2(A_{\rm T}-\lambda)\sin{2\beta} \\
\frac{\partial\Delta_{\rm sol}}{\partial\beta} & = & \frac{n_{\rm sw}}{{\rm D}_{\rm 0}|\Vec{r}(E)|f^2} \cdot \frac{\sin\beta\cos(A_{\rm T}-\lambda)(\sin\theta-\theta\cos\theta)}{\sin^3\theta},
\end{eqnarray}
and
\begin{eqnarray}
\frac{\partial\Delta_{\rm R\odot}}{\partial\lambda} & = & \frac{|\Vec{r}(E)|\cos\beta\sin(\lambda-A_{\rm T})}{c} \\
\frac{\partial\Delta_{\rm p}}{\partial\lambda} & = & \frac{\varpi|\Vec{r}(E)|^2}{2c}\cos^2\beta\sin(2\lambda-2A_{\rm T}) \\
\frac{\partial\Delta_{\rm sol}}{\partial\lambda} & = & \frac{n_{\rm sw}}{{\rm D}_{\rm 0}|\Vec{r}(E)|f^2} \cdot \frac{\sin(\lambda - A_{\rm T})\cos\beta(\sin\theta-\theta\cos\theta)}{\sin^3\theta}.
\end{eqnarray}

The Hessian matrix of the fitted parameters ($\delta N_{\rm 0}$, $\delta\varpi$, $\delta n_{\rm sw}$, $\delta\beta$, $\delta\lambda$), is then written as
\begin{equation}
    H=\sum_{i}
    \begin{pmatrix}
    \displaystyle \frac{1}{\nu^2} & \displaystyle -\frac{P_i}{\nu} &  \displaystyle -\frac{Q_i}{\nu} & \displaystyle -\frac{R_i}{\nu} & \displaystyle -\frac{S_i}{\nu} \\
\displaystyle -\frac{P_i}{\nu} & P^2_i & P_iQ_i & P_iR_i & P_iS_i \\
\displaystyle -\frac{Q_i}{\nu} & P_iQ_i & Q_i^2 & Q_iR_i & Q_iS_i \\
\displaystyle -\frac{R_i}{\nu} & R_iP_i & R_iQ_i & R_i^2 & R_iS_i \\
\displaystyle -\frac{S_i}{\nu} & S_iP_i & S_iQ_i & S_iR_i & S_i^2 \\
\end{pmatrix}
\end{equation}
where $\displaystyle\sum_{i}$ denotes summation over all TOAs. Presume the data has sufficient orbital coverage over the Earth orbit, for a given function $\mathcal{F}$ we can have $\displaystyle \sum_{i}\mathcal{F}_{i}\simeq N\cdot\overline{\mathcal{F}}$, where $N$ is the number of TOAs and the overline represents average over the course of an Earth's orbit. The covariance matrix of the fitted parameters is then written as
\begin{equation}  
    C=\overline{H}^{-1}=\begin{pmatrix}
    \displaystyle \frac{1}{\nu^2} & \displaystyle -\frac{\overline{P}}{\nu} &  \displaystyle -\frac{\overline{Q}}{\nu} & \displaystyle -\frac{\overline{R}}{\nu} & \displaystyle -\frac{\overline{S}}{\nu}\\
\displaystyle -\frac{\overline{P}}{\nu} & \overline{P^2} & \overline{PQ} & \overline{PR} &  \overline{PS} \\
\displaystyle -\frac{\overline{Q}}{\nu} & \overline{PQ} & \overline{Q^2} & \overline{QR} & \overline{QS} \\
\displaystyle -\frac{\overline{R}}{\nu} & \overline{RP} & \overline{RQ} & \overline{R^2} & \overline{RS}\\
\displaystyle -\frac{\overline{S}}{\nu} & \overline{SP} & \overline{SQ} & \overline{SR} & \overline{S^2}\\
\end{pmatrix}^{-1}.
\end{equation}
Thus, the correlation coefficients of $\delta n_{\rm sw}$ and $\delta\varpi$, $\delta\lambda$, $\delta\beta$, respectively, are given by
\begin{eqnarray} 
    \rho_{\delta\varpi,\delta n_{\rm sw}}&=&\frac{C_{23}}{\sqrt{C_{22}C_{33}}}, \label{eq:cc_posi} \\
    \rho_{\delta\lambda,\delta n_{\rm sw}}&=&\frac{C_{34}}{\sqrt{C_{33}C_{44}}}, \label{eq:cc_posi_lamda} \\
    \rho_{\delta\beta,\delta n_{\rm sw}}&=&\frac{C_{35}}{\sqrt{C_{33}C_{55}}}. \label{eq:cc_posi_beta}
\end{eqnarray}

If only considering the two delay terms caused by parallax and solar wind, the derivative timing formula is then written as
\begin{equation}
    R = -\delta N_{\rm 0} / \nu + P \cdot \delta\varpi + Q \cdot \delta n_{\rm sw}.
\end{equation}
Accordingly, the covariance matrix in this case is
\begin{equation} \label{eq:cc_noposi}
    C'=\overline{H'}^{-1}=\begin{pmatrix}
    \displaystyle \frac{1}{\nu^2} & \displaystyle -\frac{\overline{P}}{\nu} &  \displaystyle -\frac{\overline{Q}}{\nu} \\
\displaystyle -\frac{\overline{P}}{\nu} & \overline{P^2} & \overline{PQ} \\
\displaystyle -\frac{\overline{Q}}{\nu} & \overline{PQ} & \overline{Q^2} \\
\end{pmatrix}^{-1},
\end{equation}
from which gives correlation coefficient of $\delta\varpi$ and $\delta n_{\rm sw}$, is then
\begin{eqnarray} \label{eq:cc_cov_nopm}
    \rho'_{\delta\varpi,\delta n_{\rm sw}}&=&\frac{C'_{23}}{\sqrt{C'_{22}C'_{33}}} \nonumber \\
    &=&\frac{\overline{H'}_{21}\overline{H'}_{13}-\overline{H'}_{23}\overline{H'}_{11}}{\sqrt{(\overline{H'}_{33}\overline{H'}_{11}-\overline{H'}_{31}\overline{H'}_{13})(\overline{H'}_{22}\overline{H'}_{11}-\overline{H'}_{21}\overline{H'}_{12})}} \nonumber \\
    &=&\frac{\overline{P}\cdot\overline{Q}-\overline{P\cdot Q}}{\sqrt{\overline{P^2} - (\overline{P})^2} \sqrt{\overline{Q^2} - (\overline{Q})^2}}
\end{eqnarray}


\bsp	
\label{lastpage}
\end{document}